\documentclass{aa}  

\usepackage{graphicx}
\usepackage[colorlinks=true,linkcolor=blue,citecolor=blue, urlcolor=blue]{hyperref}
\usepackage[dvipsnames]{xcolor}     

\usepackage{txfonts}
\usepackage{esvect}
\usepackage{booktabs}
\usepackage{ragged2e}
\begin{document}

   \title{A slow spin to win: The gradual kinematic evolution across metallicities of the proto-Galaxy to the high-$\alpha$ disc}
   \titlerunning{A slow spin to win}


   \author{Akshara Viswanathan\inst{\ref{kapteyn},\ref{uvic}}\thanks{Corresponding author \email{astroakshara97@gmail.com}}
          \and
          Danny Horta\inst{\ref{cca},\ref{edin}}
          \and
          Adrian~M.~Price-Whelan\inst{\ref{cca}}
          \and
          Else Starkenburg\inst{\ref{kapteyn}}
          }

   \institute{Kapteyn Astronomical Institute, University of Groningen, Landleven 12, 9747 AD Groningen, The Netherlands\label{kapteyn}
         \and
         Department of Physics and Astronomy, University of Victoria, 3800 Finnerty Road, Victoria, BC, V8P 1A1, Canada\label{uvic}
         \and
             Center for Computational Astrophysics, Flatiron Institute, 162 5th Ave., New York, NY 10010, USA\label{cca}
             \and
             Institute for Astronomy, University of Edinburgh, Royal Observatory Edinburgh, Blackford Hill, Edinburgh EH93HJ, UK\label{edin}
             }

   \date{Received 19 November 2024; accepted 23 June 2025}

 
  \abstract
   {Observational studies are identifying stars thought to be remnants from the earliest stages of the Milky Way's hierarchical mass assembly, referred to as the proto-Galaxy.}
   {We used red giant stars with kinematics from \textsl{Gaia} DR3 RVS data and [$\alpha$/M] and [M/H] estimates from low-resolution \textsl{Gaia} XP spectra to investigate the relationship between azimuthal velocity and metallicity. Our aim is to understand the transition from a chaotic proto-Galaxy to a well-ordered rotating (high-$\alpha$) disc-like population.}
   {To analyze the structure of the data in [M/H]–v$_\phi$ space for both high- and low-$\alpha$ samples with carefully defined $\alpha$-separation, we developed a model with two Gaussian components in v$_\phi$, one representing a disc-like population and the other a halo-like population. This model is designed to capture the conditional distribution P(v$_\phi \mid$ [M/H]) with a two-component Gaussian mixture model with fixed means and standard deviations in the azimuthal velocities. To quantify the spin-up of the high-$\alpha$ disc population, we extended this two-component model by allowing the mean velocity and velocity dispersion to vary between the spline knots across the metallicity range used. We also compared our findings with existing literature using traditional Gaussian mixture modelling in bins of [M/H] and investigated using orbital circularity instead of azimuthal velocity.}
   {Our findings show that the metal-poor high-$\alpha$ disc gradually spins up across [M/H] $\sim -1.7$ to $-1$, while the low-$\alpha$ sample exhibits a sharp transition at [M/H] $\approx -1$. This latter result is due to the accreted (mostly \textit{Gaia}-Enceladus-Sausage) debris dominating the metal-poor end, underscoring the critical role of [$\alpha$/M] selection in studying the Milky Way’s (old, high-$\alpha$) disc evolution.}
   {These results indicate that the proto-Galaxy underwent a slow, monotonic spin-up phase over increasing metallicities rather than a rapid, dramatic spin-up at [M/H] $\sim -1$, as previously inferred in the literature.}
   {}


   \maketitle
%

\section{Introduction}\label{1}

A key objective in Galactic astronomy is to construct a cohesive formation history of the Milky Way down to its earliest times. 
More broadly, we aim to understand the extent to which a galaxy retains its formation history over cosmic timescales and to investigate the physical processes shaping disc galaxies using the Milky Way as an example (i.e. Galactic archaeology). 
The Milky Way is the perfect cosmic backyard for this because it offers detailed unparalleled, star-by-star 6D kinematics and chemical information.

The rotationally supported disc contains most of the stars in the Galaxy. 
It is well established that the disc consists of multiple components or populations, distinguished by their chemistry, kinematics, spatial extent, and age \citep{1985norris,1983gilmore,2000chiba,2010nissen,2012bovy,2013haywood,2015hayden}. 
Structurally, there is evidence of thin and thick discs with different scale heights. 
Chemically, there is a distinct dichotomy and bimodality in the abundance of $\alpha$-elements relative to iron ([$\alpha$/Fe]); the Milky Way disc manifests a high-$\alpha$ and low-$\alpha$ sequence at fixed [Fe/H] \citep{2015hayden,2023Imig}. 
Generally, the high-$\alpha$ population is older and has a larger scale height than the low-$\alpha$ population. 
While the connection between high-$\alpha$ and thick disc and low-$\alpha$ and thin disc is fairly strong in the solar vicinity, this connection weakens significantly at larger galactocentric distances \citep[e.g.][]{2015hayden}.

On the other hand, the Galactic halo is crucial for understanding the formation and evolution of our Galaxy, as it holds the remnants of ancient cosmic events and provides a window into the processes that shaped the Milky Way.
Our current understanding of Galactic stellar halo formation suggests a dual process: (i) gas accretion from cosmic filaments that drives secular evolution and  in situ star formation, and (ii) the accretion of various mass building blocks, which contribute their baryonic material and dark matter to the larger host galaxy, adding to its mass as these building blocks are consumed \citep[see recent reviews by][]{2020helmi,2024deason}.
Thanks to the combination of \textsl{Gaia} astrometry and extensive spectroscopic surveys, our understanding of stars on halo-like orbits has advanced significantly in  recent years.
A notable finding was the discovery of stars with chemistry identical to the high-$\alpha$ disc but with halo-like orbits \citep[e.g.][]{2017bonaca,2018koppelman,2018helmi,2020belokurov}. 
One interpretation of these stars is that they were born in situ and later dynamically heated to halo-like orbits, as some simulations predicted \citep[e.g.][]{2020grand}. 
The age distribution of these in situ halo or hot thick disc stars is cut off at lookback times of 8 to 11 Gyr \citep{2019gallart,2022xiang}, suggesting that the heating event, possibly a merger, occurred at z$\sim$1 to 2 \citep[see also][]{2021montalban}.
Numerous other accreted systems have also been identified as part of the accreted stellar halo \citep{1994ibata,1999helmi,2018belokurov,2018helmi,2019koppelman,2019myeong,2020yuan,2021horta,2024viswanathangiants}. 
The most significant of these is the \textsl{Gaia}-Enceladus-Sausage (GES) system, which contributes most of the accreted halo within $\sim6-30$ kpc of the Galactic centre \citep{2020naidu}. 
There is also evidence for another major building block in the inner Galaxy (\textsl{Heracles}: \citealt{2021horta}), whose stellar mass could possibly have been as massive as the GES \citep{2024bhorta}. 
It has been speculated that there may be a link between \textsl{Heracles} and a grouping of globular cluster populations classified in terms of their orbital and/or age-metallicity properties \citep{2019massari,2019kruijssen,2020forbes}. 
Based on the chemical-dynamical properties of this population and a comparison of these with cosmological simulations such as FIRE \citep{2024horta}, it is likely that \textsl{Heracles} coalesced with the primordial Galaxy before the GES ($z\gtrsim2$). 
However, deciphering if such populations arise from one single building block or multiple blocks is still an open question.

While the characterisation and origin of the main disc and halo populations are becoming clearer, the very earliest epochs of the Galaxy and the emergence of the disc remain poorly understood.
The physical mechanism behind disc formation remains a subject of active debate.
Theoretically, disc galaxies are thought to form through the dissipative collapse of gas within dark matter haloes, coupled with the conservation of angular momentum imparted during the early stages of large-scale structure formation \citep{1962eggen,1969peebles,1978white,1980fall,1987ryden}. Observationally, deep imaging and spectroscopy at high redshift (e.g. $z$ > 1–2) have revealed a population of rotationally supported, gas-rich discs already in place within the first few billion years of cosmic history \citep[e.g.][]{2009forsterschreiber,2015wisnioski,2019ubler}, supporting the view that disc formation is an early and widespread outcome of galaxy evolution. In the context of the Milky Way, recent studies discover very and extremely metal-poor stars (VMP, [Fe/H]<-2.0, and EMP, [Fe/H]<-3.0) on disc-like orbits \citep{2019sestito,2020sestito,2020dimatteo,2024viswanathan,2025viswanathan}, driving the important question of  when and how the Milky Way's old disc formed.
While there is a complex relation between the metallicity of a star and its birth time, typically metal-poor stars in the Milky Way are old, and therefore studying them can provide insights into the early epochs.
The orbits and abundance distributions of old metal-poor stars today reflect the earliest phases of the Milky Way's star formation and enrichment history \citep{2005beers,2015frebel,2018starkenburg,2019lucey,2021horta,2024ardernarentsen,2024horta,2024mccluskey}.
\citet{2022belokurov} used [Al/Fe] from \textit{APOGEE} \citep{2017majewski} to distinguish between in situ and accreted stars, identifying in situ stars down to [Fe/H] $\sim-1.5$. 
They found that the average tangential velocity of the in situ stars increased rapidly at [Fe/H] $\sim-1$. They interpreted this transition as the epoch when the Galaxy transitioned from a relatively disordered state to well-ordered rotation.
\citet{2022conroy} used abundances and ages from the H3 Survey, finding that this transition coincides with a non-monotonic rise in [$\alpha$/Fe] abundances. 
This suggests a near-instantaneous change in star formation efficiency or gas inflow \citep[see also][]{2023chen}.
At the metal-poor end, \citet{2022rix} revealed a significant concentration of metal-poor stars ([M/H]$<-1.5$) near the Galactic centre, corroborating earlier studies of the very metal-poor component in the Galactic bulge region \citep[e.g.][]{2019lucey,2020arentsen}. 
This also corroborates with other works such as \citet{2010tumlinson,2017starkenburg,2018elbadry,2021horta,2024ardernarentsen}.
The exact relationship between the proto-Galaxy and the metal-poor stars within the solar radius is still unclear, although \citet{2023belokurov} provide evidence that the proto-Galaxy's density follows a steep power law with galactocentric radius.
\citet{2024horta} used Milky Way analogues from high-resolution FIRE simulations, showing that the proto-Galactic populations in all their Milky Way analogues exhibit weak net rotation aligned with the present-day disc.
\citet{2024semenov} used the Illustris TNG50 simulations to interpret evidence of the Milky Way's early disc formation. 
They suggest that rapid early mass growth was key to early disc formation. 
However, pin-pointing the exact driver is difficult, as both hot halo formation and gravitational potential steepening occur alongside disc emergence, indicating that a concentrated potential might be a result, not a cause, of disc formation \citep{2024dillamorespin}.

Recently, \citet{2023zhang} analyzed the \textsl{Gaia} DR3 RVS kinematic sample of the metal-poor population from \citet{2023andrae} using XGBoost algorithm on \textsl{Gaia} XP spectra in the Milky Way and identified two kinematic groups in velocity–[M/H] space: one a stationary accreted halo and the other with a net prograde rotation of 80 km s$^{-1}$ associating it with the proto-Galaxy.
\citet{2023chandra} used a similar \textsl{Gaia} RVS+XP sample with an additional $\alpha$-separation to study the Milky Way evolution in a similar parameter space (orbital circularity–[Fe/H]) instead of velocities ([V$_r$,V$_\phi$,V$_z$]–[Fe/H]). 
They found that the old high-$\alpha$ disc begins at [Fe/H]$\sim$-1 dex, with a dramatic transition from a proto-Galaxy to old disc.

For this work, we built on these studies in more detail and set out to study the transition of a proto-Galactic population to a high-$\alpha$ old disc using kinematics from \textsl{Gaia} DR3 RVS sample cross-matched with \textsl{Gaia} XP spectra [$\alpha$/M] and [M/H] abundances from \citet{2023andrae} and \citet{2024li}.
Section \ref{2} presents the underlying data used in this work, and  introduces the $\alpha$-separation.
In section \ref{3} we introduce the azimuthal velocity versus metallicity space where we study the evolution of the high-$\alpha$ disc. We also compare the results from this space with \textit{APOGEE} chemical abundances to see if our inferences hold well with high-resolution spectroscopy. 
Section \ref{3.3} presents a simple two-component Gaussian mixture model (GMM) corresponding to a disc-like and halo-like population using knots across the metallicity range to carry information between different metallicities, to interpret the transition between the proto-Galactic component to a rotation-dominated high-$\alpha$ disc.
In section \ref{4.1} we repeat our analysis using orbital circularity instead of azimuthal velocity, as circularity is position independent.
In section \ref{4} we discuss the conclusions derived in this work in the context of what we already know about the proto-Milky Way and the old high-$\alpha$ disc. We also compare our results with the recent literature and discuss some limitations and future prospects in this regard.
Section \ref{5} presents a summary and our conclusions.

\section{\textsl{Gaia} DR3 XP+RVS Data}\label{2}

The \textsl{Gaia} DR3 catalogue provides low-resolution spectroscopy (XP) for approximately 200 million stars brighter than G = 17.65, with radial velocities (RVS) available for a subset of around 30 million \citep{2023gaia,2023deangeli,2023katz}. 
Subsequent studies have derived precise metallicity and $\alpha$-abundance estimates from these spectra \citep{2023andrae,2023zhanggreen,2023martin,2024li}. 
This dataset is exceptional due to its immense size, homogeneity, all-sky coverage, and relatively straightforward selection function. 
It is therefore an ideal resource to comprehensively investigate our Galaxy’s formation and evolution.

\citet{2023andrae} developed a powerful tool using a machine learning model called \textsl{XGBoost} – to estimate a star's metallicity ([M/H]) from its low-resolution XP spectrum. 
They trained this model using a robust dataset from the Apache Point Observatory Galactic Evolution Experiment \citep[\textsl{APOGEE}:][]{2017majewski}, further enhanced with metal-poor stars from \citet{2022li}. 
The metallicities used for training are highly accurate, having been validated against established spectroscopic surveys. 
For more details on how they chose specific features and validated their catalogue, we refer directly to \citet{2023andrae}. 
We leverage their metallicity data to create a refined sample of red giant branch (RGB) stars with line-of-sight velocities from \textsl{Gaia} DR3 radial velocity spectrometer (RVS). 
The following selection is made to the \citet{2023andrae} catalogue similar to their vetted RGB sample released as Table 2 in their work:

\begin{itemize}
    \item We focused on bright stars (\texttt{phot\_g\_mean\_mag < 16}) with highly reliable parallaxes (\texttt{$\varpi$/$\sigma_\varpi$ > 5}). This ensures sufficient data quality for robust metallicity estimates. 
    \item We excluded hot stars (\texttt{teff\_xgboost > 5200 K}) as their measured metallicities can be misleadingly low.
    \item We applied a series of cuts based on colour and broad-band magnitudes (\texttt{logg\_xgboost, M$_{W1}$, G-W2, G$_{BP}$-W1}) to select genuine red giants on the Hertzsprung-Russell (HR) diagram. These are listed here:
    \begin{itemize}
        \item \texttt{logg\_xgboost < 3.5}
        \item \texttt{M$_{W1}$ > -0.3 - 0.006 $\times$ (5500 - teff\_xgboost)}
        \item \texttt{M$_{W1}$ > -0.01 $\times$ (5300 - teff\_xgboost)}, where \texttt{M$_{W1}$ = W1 + 5 $\times$ log$_{10}$($\pi$/100)}
        \item \texttt{(G - W2) < 0.2 + 0.77 $\times$ (G$_{BP}$ - W1)}
    \end{itemize}
    \item We removed stars with a high probability (\texttt{>90\%}) of belonging to globular clusters (GCs). This probability comes from a catalogue by \citet{2021vasiliev}. 
\end{itemize}

This resulted in a sample of approximately 11 million RGB stars with \textsl{Gaia} XP metallicities and RVS radial velocities.

\citet{2024li} created a catalogue of alpha-over-metallicity abundance ([$\alpha$/M]) values derived from \textsl{Gaia} XP spectra using a neural network model. 
This model learns to use XP spectra to predict stellar labels but also uses and predicts the high-resolution \textit{APOGEE} spectra that lead to the same stellar labels.
This catalogue has also been cross-checked against existing surveys for accuracy, demonstrating a median error of only $\sim$ 0.05 dex in [$\alpha$/M] for stars within our sample. 
We cross-matched our clean RGB sample with this catalogue, yielding a final sample of 9,645,972 RGB stars with metallicities and $\alpha$-abundances from \textsl{Gaia} XP spectra, and full 6D phase space coordinates from \textsl{Gaia} astrometry and RVS line-of-sight velocities.

\begin{figure}
    \centering
    \includegraphics[width=\columnwidth]{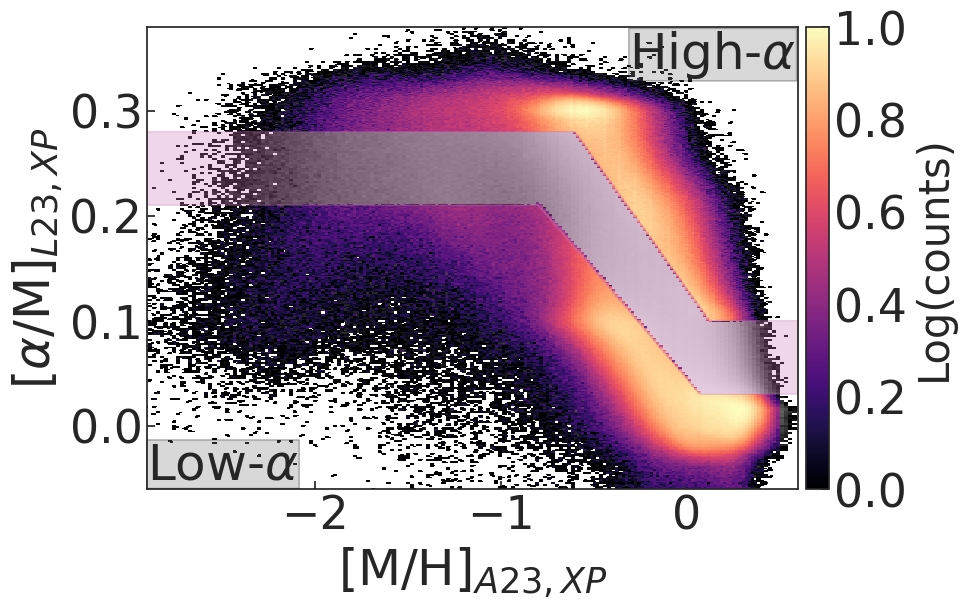}
    \caption{Logarithmic density of [$\alpha$/M] vs [M/H]. The purple band represents the high- and low-$\alpha$ sequence separation defined in this work (see text for details). Stars in the purple band are excluded. The bulk of accreted last major merger (GES) is primarily restricted to the low-$\alpha$ population with our selection.}
    \label{fig:tinsley}
\end{figure}

Figure \ref{fig:tinsley} shows the logarithmic density distribution of [$\alpha$/M] versus [M/H] of our final sample. 
While Figure \ref{fig:tinsley} showcases a clear separation between high-$\alpha$ and low-$\alpha$ stars in terms of their chemical abundance, this distinction does not necessarily translate directly to their height above the Milky Way's midplane. 
The chemical difference between these populations maps more strongly onto their radial distribution (distance from the Galactic Centre) than their vertical thickness \citep{2015hayden}. 
Therefore, we leverage this chemical separation (high-$\alpha$ versus low-$\alpha$) as a tool to differentiate the two chemically distinct disc populations and then analyze their spatial and kinematic properties (positions and velocities) independently.

\subsection{High- and low-$\alpha$ sequences}\label{2.1}

\begin{figure}
    \centering
    \includegraphics[width=0.6\columnwidth]{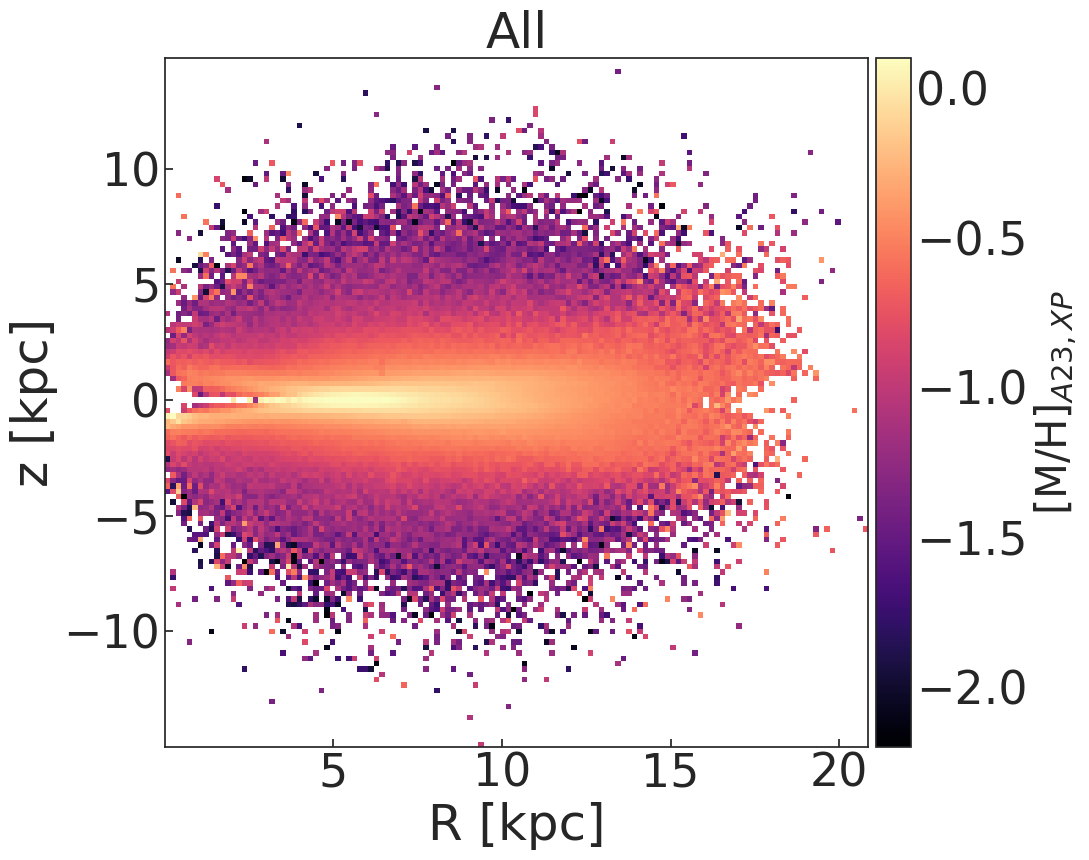}
    \includegraphics[width=\columnwidth]{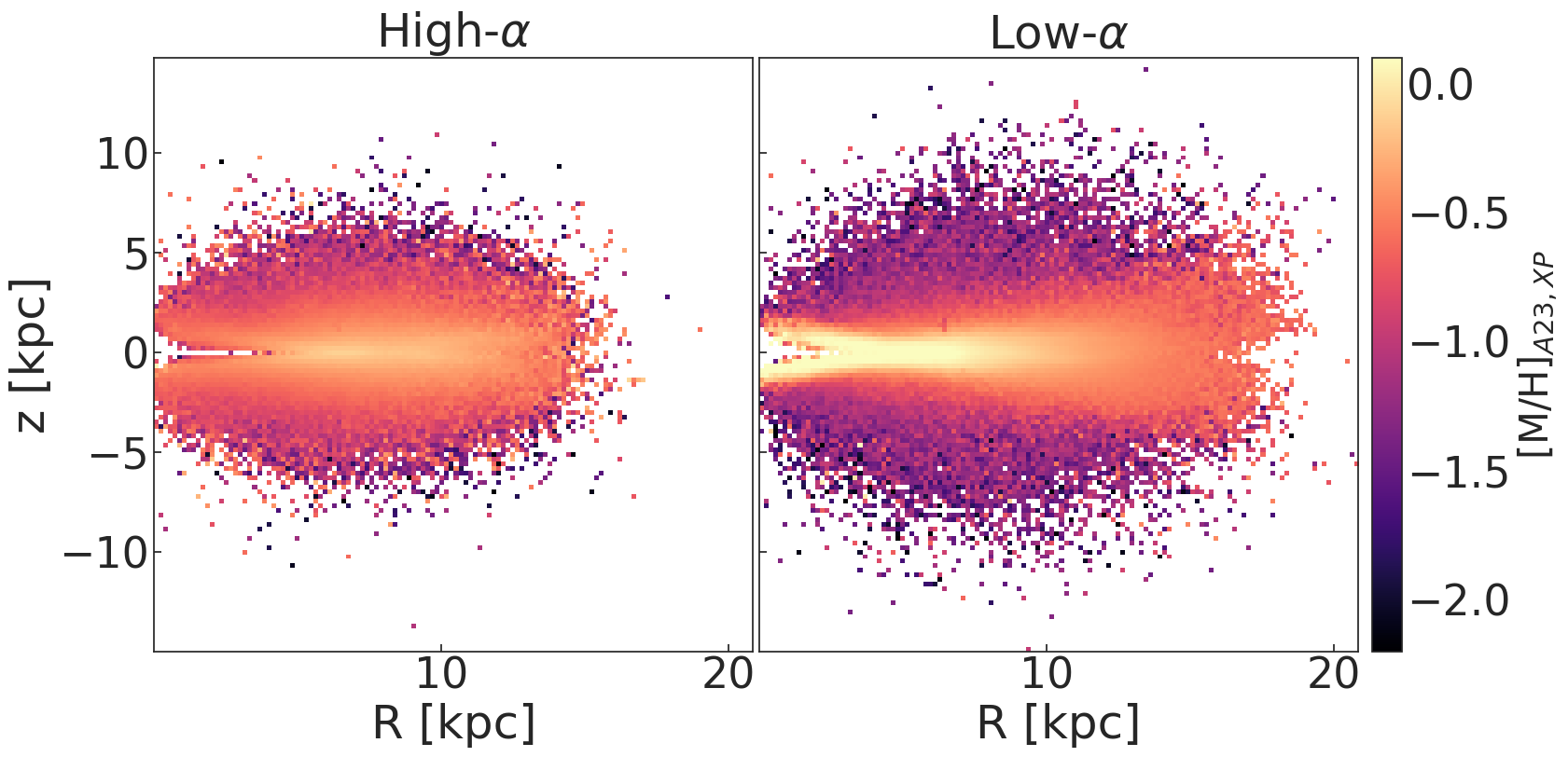}
    \caption{Distribution of stars in cylindrical Galactic coordinates (R-z plane) colour-coded by their mean metallicities for all the stars in our sample (top), high-$\alpha$ selection (bottom left), and low-$\alpha$ selection (bottom right). We can see that dust near the Milky Way's midplane significantly impacts our survey selection. We see a sharp negative metallicity gradient with respect to height above the disc plane in \textit{all stars} and low-$\alpha$ stars. High-$\alpha$ stars have a shallower negative metallicity gradient.}
    \label{fig:sky}
\end{figure}

The purple band in Figure \ref{fig:tinsley} shows the selection cut used to separate high-/low-$\alpha$ populations. 
Stars in the purple band are cut out to ensure that we have purer samples of high- and low-$\alpha$ stars. 
This is especially important for high-$\alpha$ stars since we use the high-$\alpha$ sample as the sample of stars tracing the evolution of the proto-Galaxy to high-$\alpha$ disc over metallicity and any contamination from the denser low-$\alpha$ disc can affect the purity of the high-$\alpha$ sample.
Typically, the high-alpha sequence is attributed to stars formed in situ.
However, in practise, the in situ versus accreted separation is not as simple as using an $\alpha$-separation, the consequences of which are discussed in section \ref{4.4}.

The selection is defined using the following equations:

\begin{equation}\label{eq:1}
    \rm High-\alpha=
    \begin{cases}
        \rm [M/H]<-0.6$ $\&$ $\rm [\alpha/M]>0.28\\
        \rm [M/H]\in[-0.6,0.125]$ $\&$ $\rm [\alpha/M]>-0.25\times[M/H]\\\hspace{4.9cm}+0.13\\
        \rm [M/H]>0.125$ $\&$ $\rm [\alpha/M]>0.1\\
    \end{cases},
\end{equation}

\begin{equation}\label{eq:2}
    \rm Low-\alpha=
    \begin{cases}
        \rm [M/H]<-0.8$ $\&$ $\rm [\alpha/M]<0.21\\
        \rm [M/H]\in[-0.8,0.07]$ $\&$ $\rm [\alpha/M]<-0.21\times[M/H]\\\hspace{4.9cm}+0.045\\
        \rm [M/H]>0.07$ $\&$ $\rm [\alpha/M]<0.03\\
    \end{cases}.
\end{equation}

This selection is somewhat different from the one implemented by \citet{2023chandra} who use the same sample and abundances from \textsl{Gaia} XP spectra. 
The main difference between \citet{2023chandra} and our selection are: (i) we use a more stringent selection for high-$\alpha$ to make sure that the bulk of accreted GES merger \citep{2018belokurov,2018helmi} does not contaminate the high-$\alpha$ population (see Figure \ref{fig:tinsley}), and (ii) the purple band is also quantitatively larger to ensure a sample with the highest purity. 
We tested the validity of this separation by using the $\alpha$-selection on \textsl{APOGEE} DR17 data in the [M/H]-[$\alpha$/M] space and using [Al/Fe]-[Mg/Mn] space to see if they occupy accreted, low-$\alpha$ disc or high-$\alpha$ disc stars regions as described by the tracks used in \citet{2021horta}.
The contamination rate from this validation is as low as 5\% and 8\% for high- and low-$\alpha$ stars respectively.
The high-$\alpha$ selection used here is stricter than the selection used in the literature, in order to remove as much accreted low-$\alpha$ stars as possible (mainly, GES, the last major merger). 
Even though we show the full sample in Figure \ref{fig:tinsley}, we restrict the analysis in the rest of this work to metallicities above --2.5.

\subsection{Positions and kinematics}\label{2.2}

To estimate the distance to each RGB star, we use photo-geometric distance provided by \citet{2021bailer-jones}, which is a Bayesian approach that incorporates both \textsl{Gaia} parallax measurements and a prior model of the Milky Way's stellar density distribution. 
While using a simpler method based solely on zero-point-corrected parallax \citep{2021lindegreen} yields similar results, the approach by \citet{2021bailer-jones} offers a more robust framework for estimating distances across stars with varying signal-to-noise ratios. 
To ensure that our astrometry was reliable, we focused on stars with a low renormalised unit weight error (\texttt{ruwe < 1.4}). 
This metric indicates good astrometric quality, potentially filtering out binary star systems.  
We adopted several standard assumptions: a Local Standard of Rest velocity (\texttt{V$_{LSR}$}) of 232 km s$^{-1}$, a distance of 8.2 kpc between the Sun and the Milky Way's centre \citep{2018gravity}, and the Sun's peculiar motion of (\texttt{U$_\odot$, V$_\odot$, W$_\odot$}) = (11.1, 12.24, 7.25) km s$^{-1}$ \citep{2010schonrich}. 
This allow us to calculate the positions and velocities for all the stars in our final sample.

Figure \ref{fig:sky} shows the distribution of our sample in cylindrical galactocentric radius, R, versus height above the midplane, z, colour-coded by the mean metallicity value per pixel. 
The top panel of Figure \ref{fig:sky} shows this distribution for all the stars in our sample, while bottom left and right panels show this distribution for high- and low-$\alpha$ samples, respectively. 
It is worth noting that the high-$\alpha$ stars probe a smaller range in R-z compared to low-$\alpha$ stars; most of the low-$\alpha$ stars at larger scale heights are expected to come from accretion events, and this is likely why the low-$\alpha$ sample covers a wider scale height. 
We see a strong and steep negative metallicity gradient with increasing z for the low-$\alpha$ sample; we also see this in the full sample, likely because it is dominated by stars in our low-$\alpha$ selection, while the high-$\alpha$ sample has a shallower negative metallicity gradient with increasing z. 
The high-$\alpha$ disc has a larger scale height compared to the low-$\alpha$ disc and we also see a clear disc flaring for the low-$\alpha$ disc (as seen for the Galactic thin disc) with a metallicity gradient towards larger R, reminiscent of radial migration \citep[see e.g.][]{2013haywood,2015hayden,2024bridget}.

\section{Milky Way populations in the [M/H]-v$_{\phi}$ plane}\label{3}
This section summarises the evolution of azimuthal velocity \footnote{In the rest of this paper, we use azimuthal and rotational velocities interchangeably.} versus metallicity for our high-/low-$\alpha$ samples. In examining this plane, we aim to gain an intuition of when and how quickly metal-poor populations (that for high-$\alpha$ likely contain components of the proto-Galaxy) go from a low net spin and kinematically hotter configuration to a more rotationally supported high-$\alpha$ disc population. 

\subsection{Azimuthal velocity versus metallicity trends}\label{3.1}

\begin{figure*}
    \centering
    \includegraphics[width=0.6\textwidth]{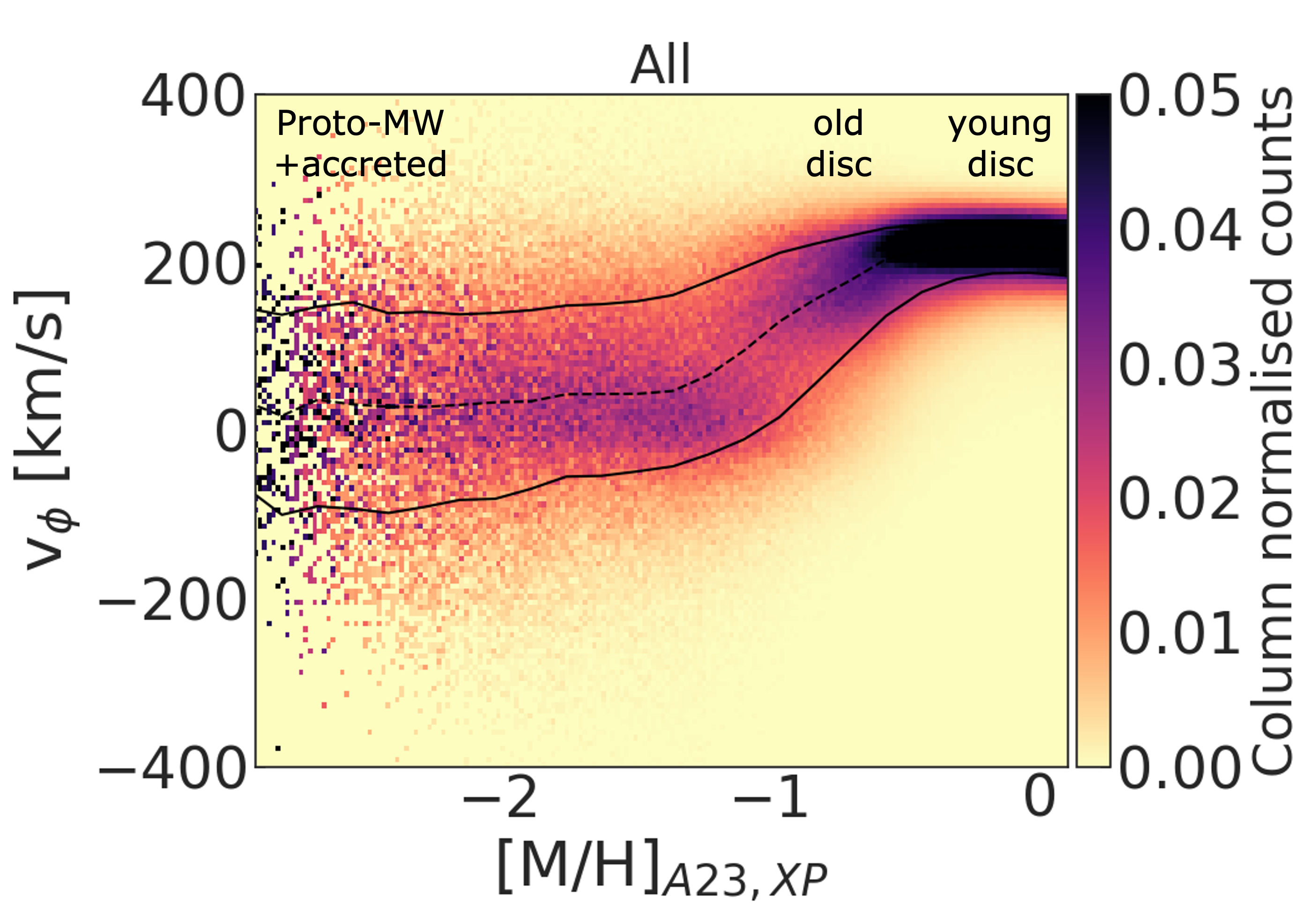}
    \includegraphics[width=\textwidth]{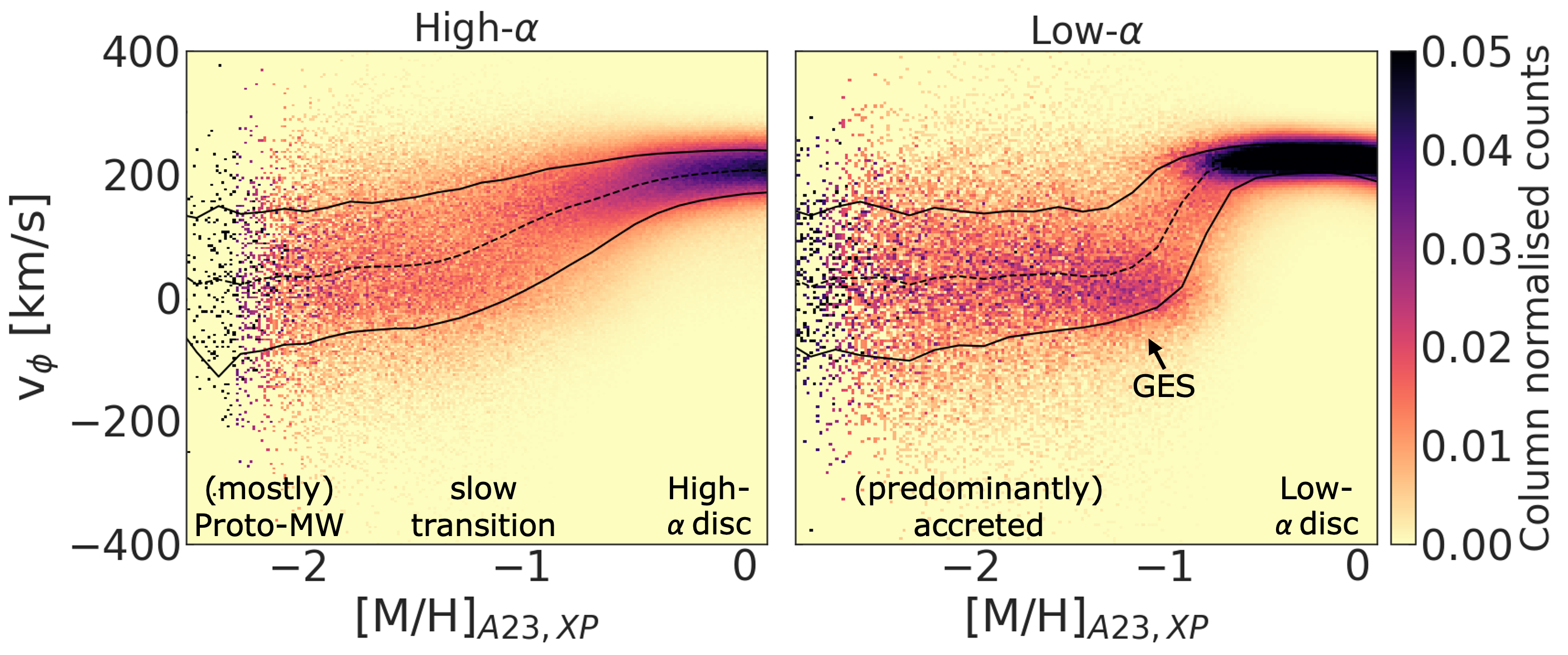}
    \caption{Column-normalised (by sum) 2D histogram of stars in the [M/H]-v$_\phi$ plane (azimuthal velocity vs metallicity) for all the stars (top), high-$\alpha$ selection (bottom left), and low-$\alpha$ selection (bottom right). The running median track is shown as dashed black line and the 16${th}$ and 84${th}$ percentile tracks are shown as black lines in all panels. The running median tracks for the low-$\alpha$ panel look more like a step-function, while high-$\alpha$ tracks is shallower, supporting the gradual monotonic increase, that can be interpreted as a gradual spin-up from old proto-Galactic populations to the present day high-$\alpha$ disc with respect to metallicities.}
    \label{fig:median-vphi-feh}
\end{figure*}

Figure \ref{fig:median-vphi-feh} illustrates the formation and evolution of the Milky Way using azimuthal velocity, v$_{\phi}$, as a function of metallicity, [M/H], for our full sample (top), and our high-/low-$\alpha$ samples (bottom). 
For all panels, each [M/H] bin has a size 0.01 dex, and we show the sum normalised histogram of azimuthal velocity, plotted as a 2D column-normalised histogram\footnote{In this work, 2D column-normalised histogram automatically means column normalisation by sum, such that the sum under the histogram curve equals 1. This is directly proportional to the probability density function of azimuthal velocity at each metallicity bin.} for all, high-, and low-$\alpha$ stars. 

Although it is intuitive to read metallicity as an age sequence, we know that the age-metallicity relation (AMR) of the Milky Way is not monotonic \citep{2014bensby,2022xiang,2024gallart,2024xiang}. 
This is true for all stars (top panel, which has a mix of in situ and accreted stars), and low-$\alpha$ stars (bottom right panel), as they are made of two different stellar populations (thin disc and accreted mergers). 
However, \citet{2022xiang} have shown that the stars in high-$\alpha$ sequence have a relatively consistent and monotonic AMR. 
They also show that the high-$\alpha$ disc reached solar metallicities at around $\sim8$ Gyr in lookback time. Therefore, the high-$\alpha$ panel should trace the first $\sim$ 5-6 Gyr of the Milky Way's evolution, which is also proposed by \citet{2023chandra} in their orbital circularity versus metallicity plane. 
All the 2D histogram in Figure \ref{fig:median-vphi-feh} have 16${th}$, 50${th}$, and 84${th}$ percentile tracks of v$_\phi$ versus [M/H] overlaid.
The median and percentile tracks look very similar if we restrict our analysis to solar neighbourhood stars (d < 3 kpc), as the velocities are less position dependent in a small volume around the Sun. 
However, in order to preserve the number statistics, especially in the low-metallicity end, we use the entire sample in the rest of this paper instead of restricting only to solar neighbourhood stars.
As an additional check, we use orbital circularity versus metallicity instead of azimuthal velocity, which is expected to be position independent in section \ref{4.1}.
We  discuss the different panels separately in the following subsections.

\subsubsection{\textit{All stars}}\label{3.1.1}

In Figure \ref{fig:median-vphi-feh} top panel, we see strongly rotational (v$_{\phi}\sim200$ km s$^{-1}$), kinematically cold thin (young, low-$\alpha$) disc dominating higher metallicities down to [M/H]$\sim$--0.7. 
Below these metallicities, we see the kinematically hotter thick (pld, high-$\alpha$) disc population (v$_{\phi}\sim180$ km s$^{-1}$) which then has a disconnection at [M/H]$\sim$--1.0. 
Furthermore, it is possible to see the GES merger (and other accretion events) dominating radial orbits (v$_{\phi}\sim0$ km s$^{-1}$) at metallicities below [M/H] $<-1.0$. 
This creates a step function behavior (also seen in the running median tracks) around the metallicities between --1.3 and --0.9 which has been reported in the literature as the spin-up phase, where we have the dramatic transition from what has been dubbed a ``proto-Galactic'' population with low net rotation to a rotation-supported disc component \citep{2022belokurov,2023chandra,2023zhang,2024kurbatov}.

\subsubsection{Low-$\alpha$ stars}\label{3.1.2}

In Figure \ref{fig:median-vphi-feh} bottom right panel, we see strongly rotation-supported low-$\alpha$ disc (v$_{\phi}\sim200$ km s$^{-1}$) dominating higher metallicities ([M/H] $>-1$) and is almost fully disjoint to the accreted population at lower metallicities ([M/H] $<-1$) that are radial and present (almost) isotropic rotational velocities (v$_{\phi}\sim0$ km s$^{-1}$). Here, we can clearly see a step function behaviour at [M/H]$\sim$--1.0, which demarks the transition from the accreted populations to the low-$\alpha$ disc. 
We see little-to-no evolution in the azimuthal velocity and velocity dispersions for the low-$\alpha$ disc stars with increasing metallicities. 
This is also because metallicity clearly does not trace age for low-$\alpha$ disc stars, but instead the birth radius of the star \citep[e.g.][]{2024bridget}, i.e. metallicities for low-$\alpha$ stars cannot be read as a temporal axis.

\subsubsection{High-$\alpha$ stars}\label{3.1.3}

In Figure \ref{fig:median-vphi-feh} bottom left panel, we see high-$\alpha$ stars with their azimuthal velocity evolving across the metallicity range monotonically. 
Both the 2D column-normalised histogram and the median tracks show the evolution of a less rotationally supported (kinematically hotter) metal-poor component (likely the proto-Galaxy), with a small but non-negligible mean azimuthal velocity value (v$_{\phi}\sim50$ km s$^{-1}$, \citealt{2024mccluskey,2024semenov,2024horta}). As metallicity increases, we see the high-$\alpha$ sample increasing in v$_{\phi}$ towards more prograde orbits, gradually transitioning from a hotter less rotating component to a rotation-dominated high-$\alpha$ disc (v$_{\phi}\sim180$ km s$^{-1}$).
The velocity dispersion for high-$\alpha$ disc orbits becomes smaller with increasing metallicities.
Therefore, Figure \ref{fig:median-vphi-feh} likely illustrates the emergence of the high-$\alpha$ disc, revealing the evolution of proto-Galactic populations gradually spinning-up with increasing metallicities. 

It is important to note that our [$\alpha$/M] selection would not fully remove accreted stars in the high-$\alpha$ sample, especially in the metal-poor end of accretion events, [M/H]<--1.3, where GES and other debris like \textsl{Heracles} \citep{2021horta} appear. 
Therefore, accreted halo stars would still be present (although probably contributing a small fraction) in the metal-poor end.
Additionally, the mean velocities and velocity dispersion that we show are present day velocity distributions.
Therefore, Figure \ref{fig:median-vphi-feh} shows the instantaneous velocity information for these populations, and not the velocity profile at formation; for these metal-poor (old) populations, these two could be drastically different due to dynamical heating over time, for example.
It is also important to note that the hotter high-$\alpha$ disc (in situ halo or hot thick disc) population, is still present in the high-$\alpha$ sample, but smaller in number than the rotation-supported high-$\alpha$ disc stars.
These "hot" high-$\alpha$ disc stars can be easily seen in row-normalised [M/H]-v$_\phi$ plane as shown in Appendix \ref{C}.

From the bottom left panel of Figure \ref{fig:median-vphi-feh}, we find evidence in support of a metal-poor high-$\alpha$ population (likely part of the proto-Galaxy) gaining rotation at a slower pace across wider range of metallicities (between -2 and -0.7) that eventually settles into a high-$\alpha$ disc population. 

\subsection{Azimuthal velocity versus metallicity tracks using high-resolution \textit{APOGEE} abundances}\label{3.2}

\begin{figure*}
    \centering
    \includegraphics[width=\textwidth]{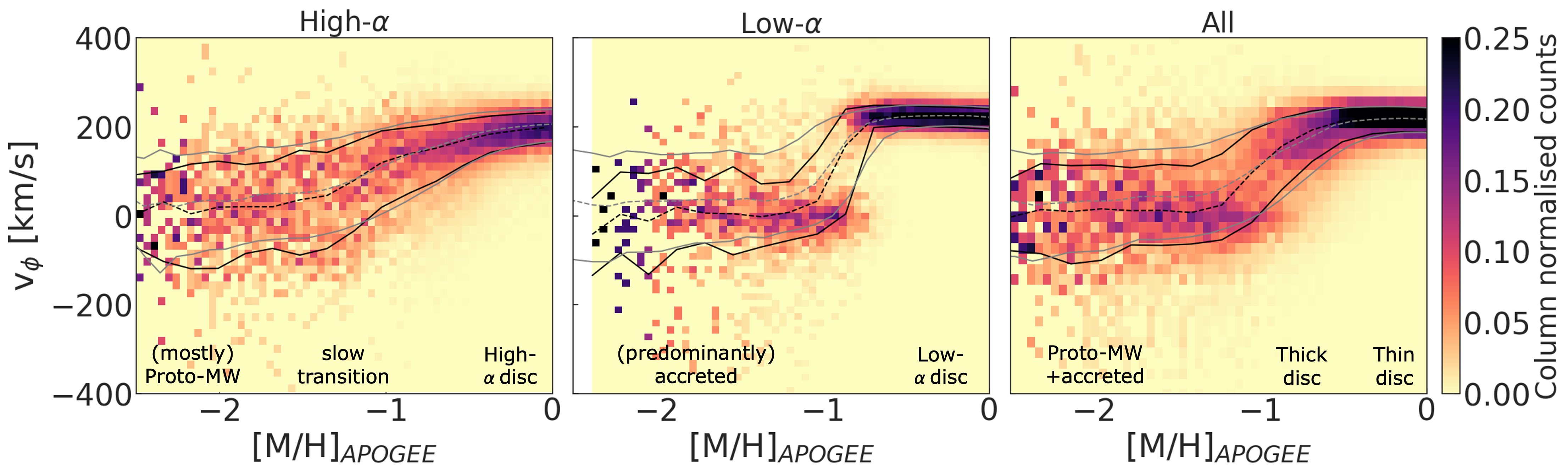}
    \caption{Column-normalised (by sum) 2D histogram of stars in the [M/H]-v$_\phi$ plane for \textit{APOGEE} DR17 giants (with the same high- and low-$\alpha$ separation as shown in Figure \ref{fig:tinsley}) for all the stars (right), high-$\alpha$ selection (left), and low-$\alpha$ selection (middle). The running median track is shown as dashed black line and the 16${th}$ and 84${th}$ percentile tracks are shown as black lines in all panels. The median, 16${th}$ and 84${th}$ percentile tracks for our \textsl{Gaia} XP sample are shown in grey for comparison. The tracks between \textsl{Gaia} XP and \textit{APOGEE} samples show remarkable resemblance in high-$\alpha$ and small differences for low-$\alpha$ and \textit{all stars} due to lower contamination in low-$\alpha$ selection in \textit{APOGEE} (see Appendix \ref{A}).}
    \label{fig:median-apogee}
\end{figure*}

Figure \ref{fig:median-apogee} shows 2D column-normalised histograms of [M/H]-v$_\phi$ plane for high-$\alpha$ (left), low-$\alpha$ (middle), and \textit{all stars} (right) for \textsl{APOGEE} DR17 stars with the same $\alpha$-selection described by equations \ref{eq:1} and \ref{eq:2}.
In \textsl{APOGEE} DR17, we select stars with \texttt{log g < 3.5}, excluding potentially problematic data based on quality flags such as \texttt{ASPCAPFLAG}, and \texttt{WARNING}, or \texttt{BAD} flags defined in ASPCAP for T$_{\rm eff}$, log \textit{g}, [M/H], and [$\alpha$/M].
The running 16${th}$, 50${th}$, and 84${th}$ percentile tracks for the \textit{APOGEE} data are shown as black lines and the tracks from our final sample in Figure \ref{fig:median-vphi-feh} are shown in grey in all three panels. 
We find very good resemblance between the track in the \textsl{APOGEE} and \textsl{Gaia} XP data, especially for high-$\alpha$ stars stars.
It showcases that the scenario of a gradual spin-up of the proto-Galaxy to the high-$\alpha$ disc population (across metallicities) is not due to large errors in \textit{Gaia} XP abundances, as it is also supported by a much more reliable high-resolution chemical abundance data from \textit{APOGEE}.
The low-$\alpha$ density distribution (and median tracks) are the ones that vary the most between the two samples.
This would be expected if contamination from in situ centrally concentrated stars from the proto-Galaxy were present in our low-$\alpha$ sample, while \textit{APOGEE} has a much cleaner and purer low-$\alpha$ selection, due to higher quality of the chemical abundances.
This can also be seen by the low-$\alpha$ disc and halo populations being completely disjoint in the low-$\alpha$ 2D histograms. 
This is discussed more in detail in Appendix \ref{A} using [M/H]-v$_\phi$ plane towards and away from the inner Galaxy and noting the differences for low-$\alpha$ 2D histograms in Figure \ref{fig:median-apogee}.
Lastly, the \textit{all stars} panel in \textsl{APOGEE} also resembles what we see in our \textsl{Gaia} XP sample, with stars at lower metallicities dominated by accreted stars and proto-Galaxy populations with low net spin, and higher metallicities dominated by the kinematically hotter high-$\alpha$ disc and kinematically colder low-$\alpha$ disc at even higher metallicities. 

Given these results, we are confident that our selection presented in Figure \ref{fig:tinsley} is efficient to separate different stellar populations in the Milky Way, where we can analyse the [M/H]-v$_\phi$ plane. 
In the following section, we set out to model these data using a new mixture model, with the aim of quantifying the point at which metal-poor populations transition into the more rotationally supported disc populations; we also aim to assess the fraction of halo/disc populations at different metallicity bins.

\section{Modelling the high-/low-$\alpha$ stars in the [M/H]-v$_{\phi}$ plane}\label{3.3}

\begin{figure*}
    \centering
    \includegraphics[width=0.39\linewidth]{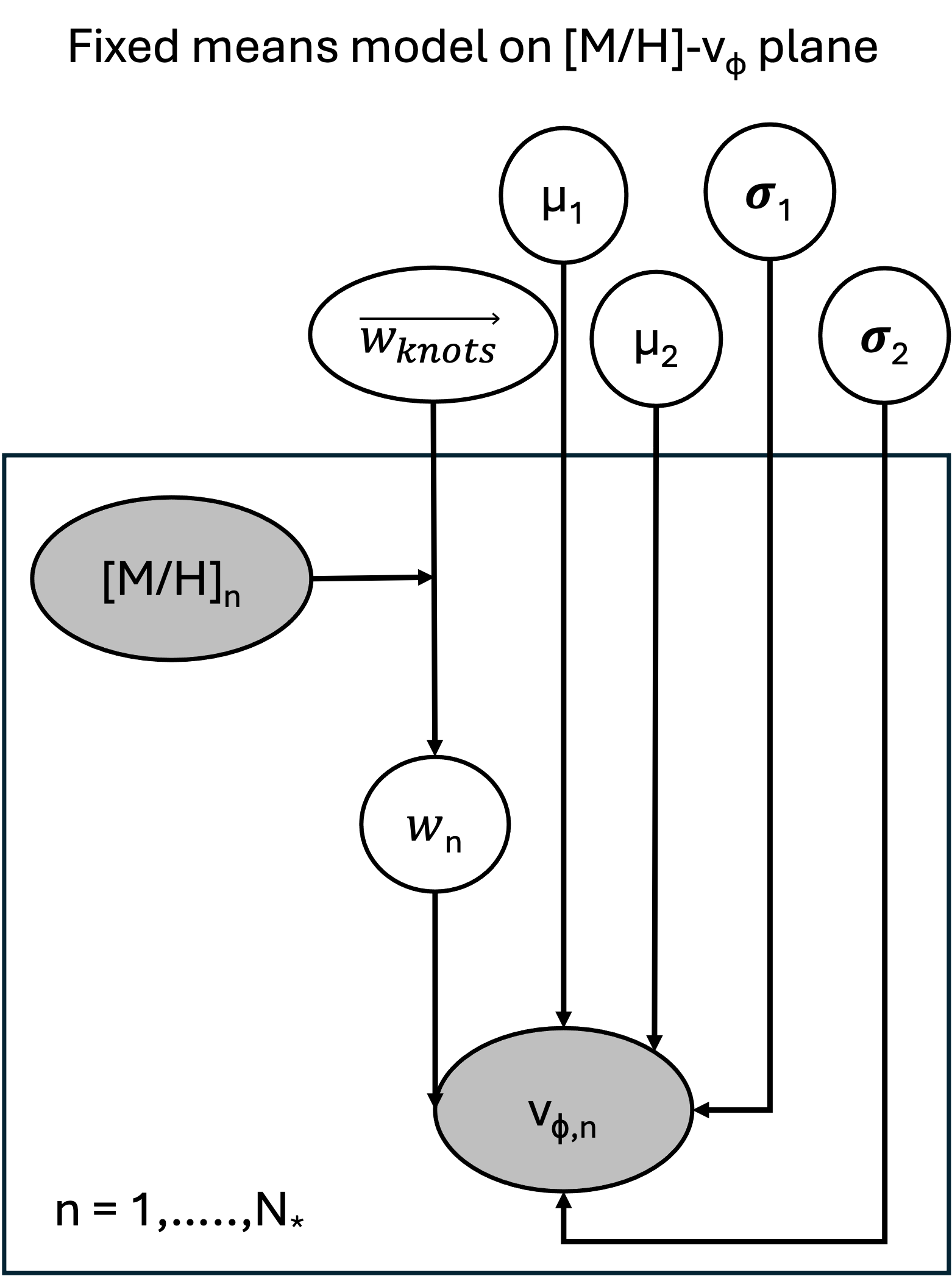}
    \includegraphics[width=0.51\linewidth]{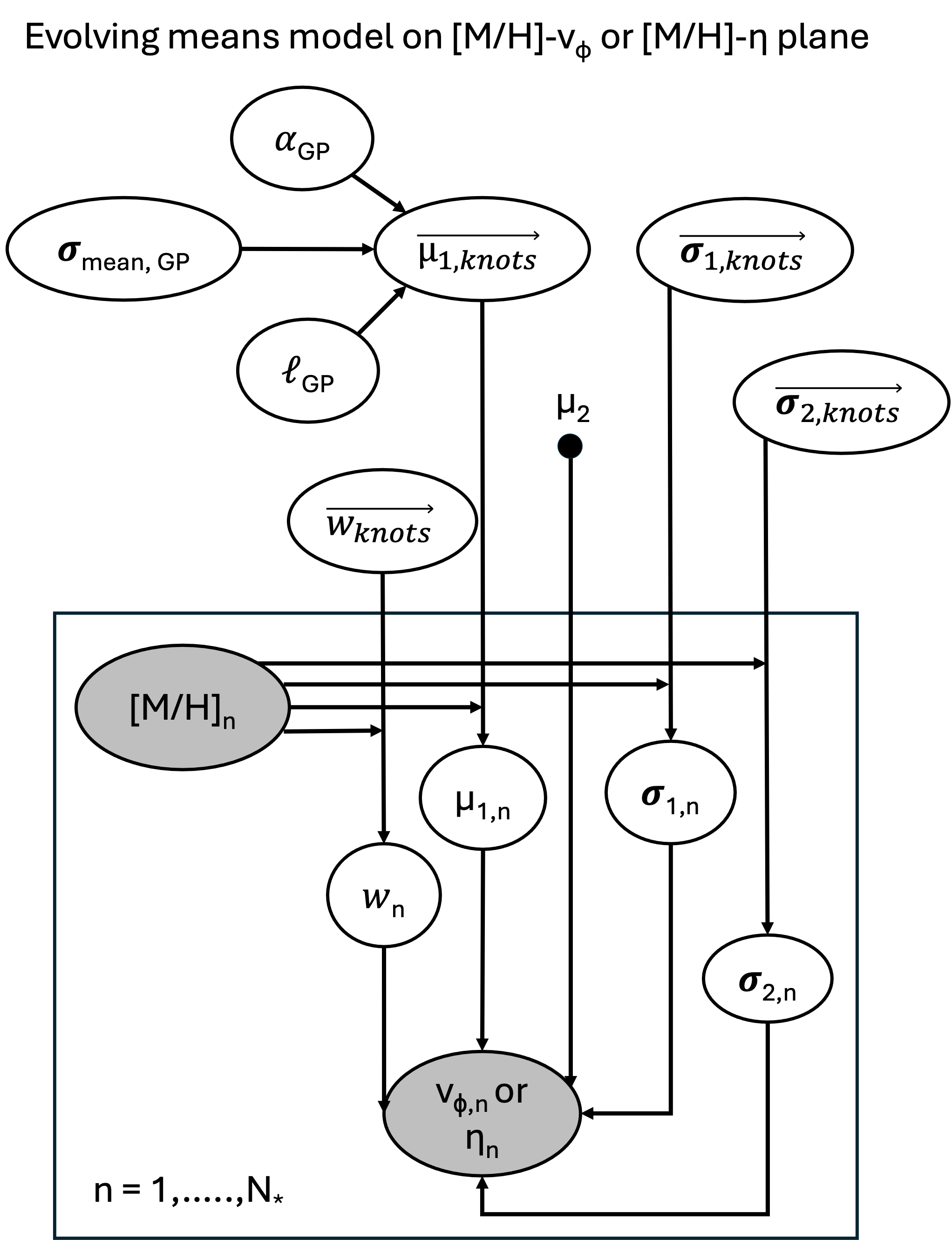}
    \caption{Graphical model representation of the evolving means model, modelling the conditional distribution P(v$_\phi$|[M/H]) with a two-component GMM, where the means and standard deviations are fixed (left) and varying (right) with respect to the underlying metallicities. The evolving or varying means model is also implemented for circularity as the conditional distribution or P($\eta$|[M/H]). The subscript 1 and 2 stands for the disc-like and halo-like components, respectively, with $\mu$ and $\sigma$ as the Gaussian mean and standard deviations in v$_\phi$ or $\eta$, conditioned on the value of [M/H]. $w$ stands for the relative contribution of the disc-like component with the relative contribution of the halo-like component defined as (1-$w$). A list of the model parameters and their priors and functional forms are given in Table \ref{func-form}.}
    \label{fig:graphic-model}
\end{figure*}

\begin{figure*}
    \centering
    \includegraphics[width=0.491\textwidth]{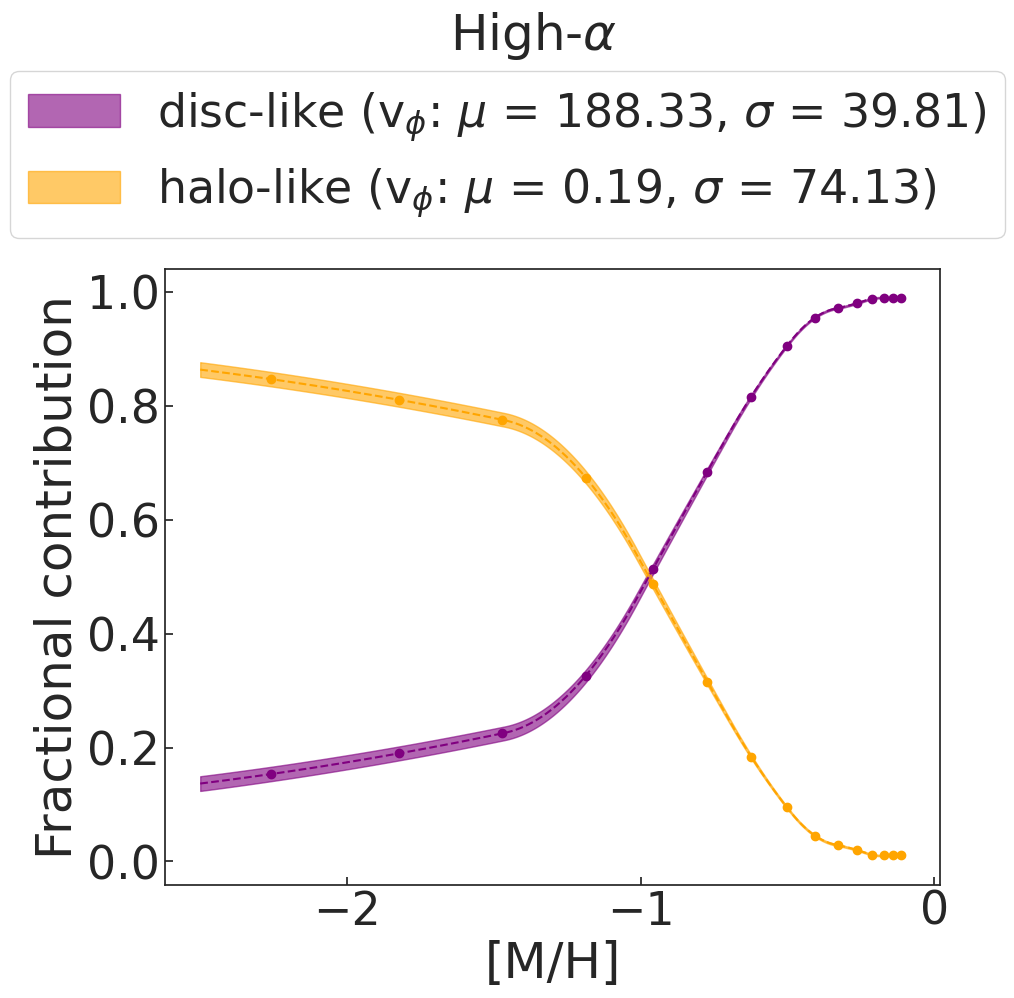}
    \includegraphics[width=0.49\textwidth]{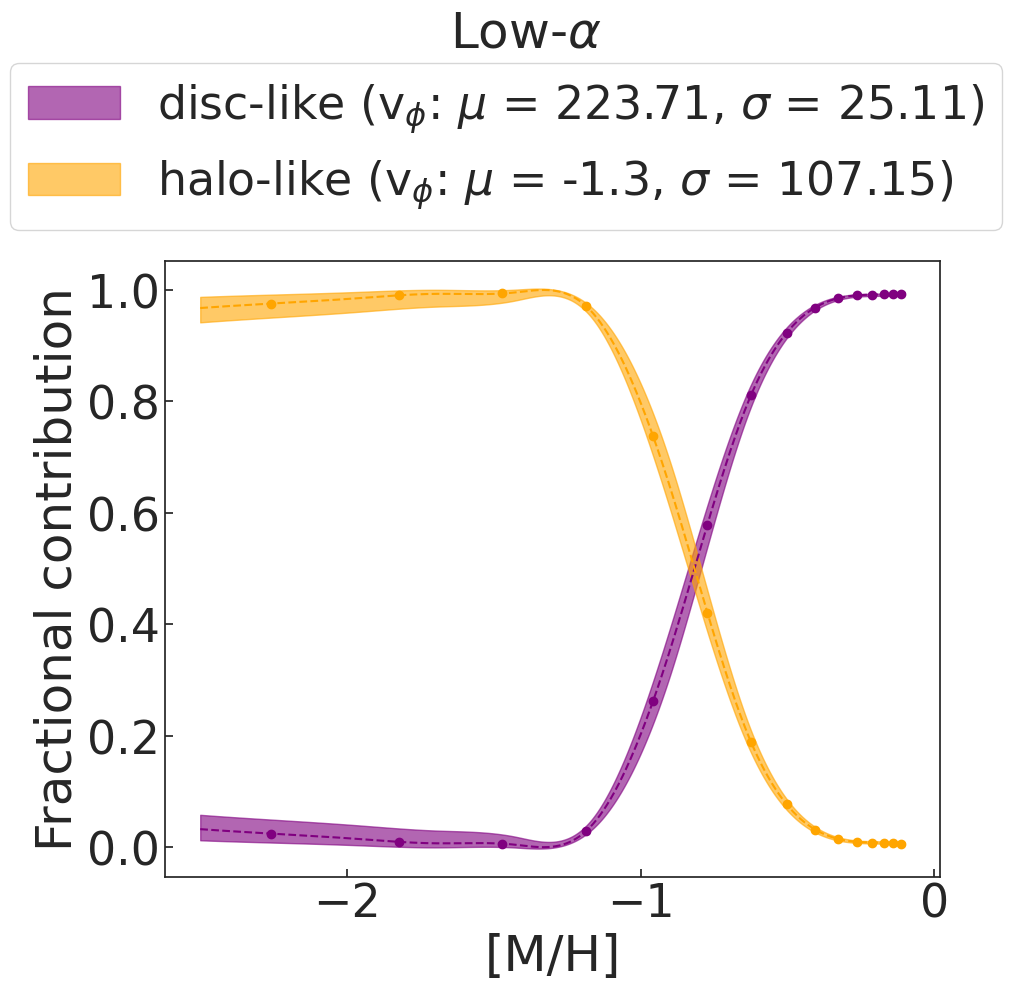}
    \caption{Fractional contribution of disc-like (purple) and halo-like (orange) GMM components as a function of metallicity for high- and low-$\alpha$ stars. The scatter points show the location of the knots chosen to run our model. The bands are $1\sigma$ uncertainties on the converged weights. These fractions are computed for a fixed Gaussian for halo and disc-like stars when the MCMC is converged. The converged azimuthal velocity and velocity dispersion is shown in the legend, shown in units of km s$^{-1}$. The low-$\alpha$ model clearly shows two separate populations (thin disc and accreted) shown by the step function that could be misclassified as a rapid spin-up, while the high-$\alpha$ model shows a shallower monotonically increasing(decreasing) profile for the disc(halo) with [M/H].}
    \label{fig:frac-model}
\end{figure*}

To decipher the underlying structure of the data in [M/H]-v$_\phi$ space for both the high-/low-$\alpha$ samples, we create a simple model described by two Gaussian components in v$_\phi$ (one that is more disc-like and other that is more halo-like) the probabilities of which is conditioned on the value of the metallicity, thereby carrying over the two-component GMM information across metallicities. 
This model is implemented using \texttt{numpyro} \citep{2019phan,2019bingham}, a lightweight probabilistic programming library that provides a \texttt{numpy} backend for \texttt{pyro}.

We argue that this model is more well-suited for this problem than a standard GMM decomposition (even if the number of components is set free and determined using the Bayesian Information Criteria) in bins of [M/H], because the underlying Gaussians are independent of each other in between different [M/H] bins. 
Therefore, instead of manually linking different Gaussians across metallicity, our model $P(v_\phi\ \mid [M/H])$ is able to connect the two Gaussians across the metallicities as a continuous conditional distribution. 
This is also advantageous as it allows us to follow and interpret the different Gaussian components and thus examine how they vary across metallicity space. An underlying assumption of our method is that we can model the low-$\alpha$ and high-$\alpha$ discs as a Gaussian distribution in v$_{\phi}$ space. To first order, this is a good approximation (see middle panels of Figure \ref{fig:circ-main}); it is also a good approximation for the stellar halo, as our sample is dominated by the debris of one major accreted satellite (i.e. GES). 

Moreover, our model allows us to interpret how different stellar populations evolve in v$_{\phi}$ across [M/H]. It also enables us to measure the transition from a halo-like non-rotating population to a disc-like rotating one for both the high-/low-$\alpha$ samples. We reason that the two-component fit is also physically motivated as we know that the Galaxy has a halo and a disc in the high- and low-$\alpha$ regime. 
While simple in model form, we show here how this model is able to accurately capture the data.

In practice, the model takes in [M/H] and v$_\phi$ as input parameters, and uses priors on two (separate) Gaussian components to split the data into a halo-like and disc-like population across [M/H]. In detail, we choose 16 evenly spaced knots on a log scale\footnote{We space the knots in the spline on a log scale to account for the fact that there are many more metal-rich stars than metal-poor stars.} across the metallicity space from $-2.5<$ [M/H] $<0.01$. For each batch of data in these bins, our model computes the fraction of the data that can be modelled by one Gaussian compared to the other (i.e. the fractional contribution of each Gaussian); this allows us to quantify how much of the data is better fitted by a halo-like populations versus a disc-like one in each [M/H] bin. We set the priors of each Gaussian component as a normal distribution with $\cal{N}(\mu, \sigma)$, where $\mu=0$ km s$^{-1}$ and $\sigma=150$ km s$^{-1}$. For the priors on the weights (or fractional contribution) of each Gaussian at the locations of each [M/H]-knot in the spline, we use a Dirichlet prior with a constant concentration value of 0.5 for both the components\footnote{We chose a Dirichlet distribution to ensure that the  sum of the two weights should always be equal to 1.0, as expected for the fractional contribution parameter.}. The means, standard deviations, and relative weights for the two Gaussian components are then sampled using a Hamiltonian Monte Carlo (HMC) inference, using the No U-Turn Sampler (NUTS). We run the sampler for 200 warm-up steps, and 100 sampling steps. This results in the Markov chain Monte Carlo fits for high-$\alpha$ and low-$\alpha$ stars in our sample. 
We show the graphical model representation of this model in the left panel of Figure \ref{fig:graphic-model} and a list of model parameters and their functional forms are provided in the second column of Table \ref{func-form}.

When running this method on our high-/low-$\alpha$ samples, we find that the \texttt{r\_hat} split Gelman Rubin diagnostic parameter is less than 1.1 for the majority of the [M/H] bins, indicating that the chains have converged. Moreover, the chains look stable, and \texttt{n\_eff}, the number of effective samples, is at least 30 or more for the means, standard deviations, and relative weights.
We do note however that for the lowest [M/H] bins, the \texttt{r\_hat} values are higher for high-$\alpha$ stars (up to 1.72 for the knots in the metal-poor regime, even though the chains are converged) than for the low-$\alpha$ stars (oscillate below 1.1 regardless of the warm up steps and sample size for the MCMC chains). We tested varying the number of warm-up and sampling steps, but found that this did not affect our results. This result indicates that the samples we generate are a fair representation of the posterior probability distribution over the model parameters for the low-$\alpha$ stars. This is also the case for the metal-rich bins in the high-$\alpha$ sample. However, at lower [M/H] for high-$\alpha$ stars, our model requires more flexibility to generate a fair representation of posterior distributions. 
The low-$\alpha$ sample is comprised of two clearly distinct populations, the low-$\alpha$ disc and GES debris at different metallicities. 
Conversely, the high-$\alpha$ sample is comprised of stars that appear to follow a single trajectory across the entire metallicity range. 
Moreover, metal-poor high-$\alpha$ populations are likely an amalgamation of stellar populations (i.e. \textsl{Heracles}, old in situ material). Thus, our results hint that our fixed means two-Gaussian components are too simple to be able to capture the distribution of this data perfectly. 
We can see the results from this model in Figure \ref{fig:median-vphi-feh}. Here, the low-$\alpha$ stars cleanly separate into two distinct stellar populations, each of which dominates a different metallicity range; we note that the median tracks display  almost a step function. At high (low) [M/H], the low-$\alpha$ disc (GES debris) dominates. This is not the case for the high-$\alpha$ star sample, that displays a monotonically increasing behaviour in v$_{\phi}$ for increasing metallicity.
Therefore, a model with the means and standard deviations of the azimuthal velocity varying across the metallicities may be more suited for the high-$\alpha$ stars. 

However, despite this limitation, it is useful to compare the results from our model for the high- and low-$\alpha$ population. Thus, we proceed with this simple two-component model for both high-/low-$\alpha$ populations to assess the fractional contribution of disc-like and halo-like components in each sample.

\subsection{Interpretation of the frozen means model in the [M/H]-v$_\phi$ on high/low-$\alpha$ stars}

Figure \ref{fig:frac-model} shows the fractional contribution of a halo-like component in orange and disc-like component in purple with the $1\sigma$ uncertainty in the fractions as orange and purple shaded bands, respectively. 
The knots in metallicity chosen to run the model are shown as scatter points. 
The converged velocity means for high-/ low-$\alpha$ stars are 188 km s$^{-1}$ and 223 km s$^{-1}$, respectively. These values match well the mean rotational velocity of the high-$\alpha$ (thick) and low-$\alpha$ (thin) discs. 
The corresponding velocity dispersions are 39 km s$^{-1}$ and 25 km s$^{-1}$ for high-/low-$\alpha$ discs, showing that the high-$\alpha$ disc is kinematically hotter than the low-$\alpha$ disc. Furthermore, the halo-like components have a more radial mean velocity and hotter velocity dispersion when compared to their respective high-/low-$\alpha$ disc samples. Here, the high-$\alpha$ halo-like component has a smaller velocity dispersion (74 km s$^{-1}$) than low-$\alpha$ stars (107 km s$^{-1}$). 
This could simply be due to the fact that the two-component fit with fixed mean velocities and standard deviations might not be the best representation of the underlying data for high-$\alpha$ stars. 
Conversely, it could also imply that the high-$\alpha$ halo component is kinematically colder than the low-$\alpha$ (GES) one.

The main difference we can see between the fractional contribution of disc-like and halo-like stars in high-/low-$\alpha$ subsamples is that the low-$\alpha$ stars completely switch between halo and disc around a very narrow range in metallicities between -1.1 and -0.8. 
This step function behaviour is clearly seen in the right panel of Figure \ref{fig:frac-model}.
The small fall in halo-like fractions at lower metallicities could simply be due to noisy data in the metal-poor end. 
For the low-$\alpha$ stars, within the uncertainties in the weights, we can clearly see that the metal-poor end is fully composed of halo-like stars ([M/H$\leq-1.1$]) and the metal-rich end is fully composed of disc-like stars ([M/H]$\geq-0.6$). Conversely for high-$\alpha$ stars, we do not see such a steep turn-over between the halo-like and disc-like samples. In contrast, the disc-like component gradually increases with increasing metallicity in its relative fraction, while the halo-like component gradually decreases.
The disc-like component is present with 18\% relative contribution in the very metal-poor end ([M/H]<--2) in this simple model.
This result favours a more gradual spin-up (with respect to metallicities) scenario. However, as discussed above, our model is not able to fully capture the distribution of the high-$\alpha$ stars in the metal-poor regime. To improve the underlying model to better capture the data in order to quantify the spin up of the high-$\alpha$ disc, we run a separate model in the following section that is able to let the mean velocity and velocity dispersion evolve over increasing metallicities. 

\subsection{Quantifying the evolution of the high-$\alpha$ disc with metallicity}\label{3.3.1}

\begin{table*}
    \caption{Model parameters, their priors, and their functional forms for the three different models presented in this work.}
    \label{func-form}
\centering
\begin{center}
\begin{tabular}{c|c|c|c}
     \toprule
     & \multicolumn{3}{c}{Prior / Functional form} \\
     \cline{2-4}
     Parameter& Frozen means model & Evolving means model & Evolving means model \\
     &on [M/H]-v$_\phi$ plane&on [M/H]-v$_\phi$ plane&on [M/H]-$\eta$ plane\\ 
     \hline &&&\\
     $\vv{w_{knots}}$   & $\rm Dirichlet(0.5 \cdot 2)$    & $\rm Dirichlet(0.5 \cdot 2)$ &   $\rm Dirichlet(0.5 \cdot 2)$\\
     $\sigma_{\rm mean, GP}$&   -  & $\mathcal{U}(50$ km s$^{-1}$, $200$ km s$^{-1}$$)$   &$\mathcal{U}(0.1, 0.9)$\\
     $\ell_{\rm GP}$ & - & $\mathcal{U}(0,1)$&  $\mathcal{U}(0,1)$\\
     $\alpha_{\rm GP}$  & - & $\mathcal{U}(0,4)$&  $\mathcal{U}(0,4)$\\
     $\vv{\mu_{1,knots}}$ &   -  & $\mathcal{N}(80$ km s$^{-1}$$, cov_{spin}$ $_{up})$ &$\mathcal{N}(0.3, cov_{spin}$ $_{up})$\\
     $\vv{\sigma_{1,knots}}$& -  & $\mathcal{H}\mathcal{N}(150$ km s$^{-1}$$)$   &$\mathcal{H}\mathcal{N}(0.3)$\\
     $\vv{\sigma_{2,knots}}$ & -  & $\mathcal{T}\mathcal{N}(100$ km s$^{-1}$$, 50$ km s$^{-1}$$;$ &$\mathcal{T}\mathcal{N}(0.4, 0.2; 0.4, 0.7)$\\
     &&$50$ km s$^{-1}$$, 150$ km s$^{-1}$$)$&\\
     $w$ & $\mathcal{S}_{interp}([M/H] \mid \vv{knots_{w}}, \vv{w_{knots}})$ & $\mathcal{S}_{interp}([M/H] \mid \vv{knots_{w}}, \vv{w_{knots}})$ & $\mathcal{S}_{interp}([M/H] \mid \vv{knots_{w}}, \vv{w_{knots}})$\\
     $\mu_1$ & $\mathcal{N}$(0 km s$^{-1}$, 150 km s$^{-1}$) & $\mathcal{S}_{interp}([M/H] \mid \vv{knots_{\mu}},\vv{\mu_{1,knots}})$ & $\mathcal{S}_{interp}([M/H] \mid \vv{knots_{\mu}},\vv{\mu_{1,knots}})$\\
     $\sigma_1$ & $\log_{10} \frac{\sigma_1}{\rm km s^{-1}}$ $\sim$ $\mathcal{U}$(0, 2.5) & $\mathcal{S}_{interp}([M/H] \mid \vv{knots_{\mu}},\vv{\sigma_{1,knots}})$ & $\mathcal{S}_{interp}([M/H] \mid \vv{knots_{\mu}},\vv{\sigma_{1,knots}})$\\
     $\mu_2$ & $\mathcal{N}$(0 km s$^{-1}$, 150 km s$^{-1}$) & Fixed, 0 km s$^{-1}$ & Fixed, 0\\
     $\sigma_2$ & $\log_{10}\frac{\sigma_2}{\rm km s^{-1}}$ $\sim$ $\mathcal{U}(0, 2.5)$ & $\mathcal{S}_{interp}([M/H] \mid \vv{knots_{\mu}},\vv{\sigma_{2,knots}})$ & $\mathcal{S}_{interp}([M/H] \mid \vv{knots_{\mu}},\vv{\sigma_{2,knots}})$\\
      & $w \cdot \mathcal{N}(\mu_1,\sigma_1) + $ & $w \cdot \mathcal{N}(\mu_1,\sigma_1) + $ & $w \cdot \mathcal{F}\mathcal{N}(\mu_1,\sigma_1; -1, 1) + $\\
     v$_\phi$ or $\eta$ & $(1-w) \cdot \mathcal{N}(\mu_2,\sigma_2)$ & $(1-w) \cdot \mathcal{N}(\mu_2,\sigma_2)$ & $(1-w) \cdot \mathcal{F}\mathcal{N}(\mu_2,\sigma_2; -1, 1)$ \\
     \bottomrule
\end{tabular}
\end{center}
\justifying
Note: Abbreviations - $\mathcal{H}\mathcal{N}$ = Half-normal distribution, $\mathcal{T}\mathcal{N}$ = Truncated normal distribution, $\mathcal{S}_{interp}$ = spline interpolation, $\mathcal{F}\mathcal{N}$ = Folded normal distribution. The parameters of $\mathcal{T}\mathcal{N}$ and $\mathcal{F}\mathcal{N}$ are ($\mu$, $\sigma$; min, max). The subscript 1 and 2 stands for the disc-like and halo-like components, respectively, with $\mu$ and $\sigma$ as the Gaussian mean and standard deviations in v$_\phi$ or $\eta$, conditioned on the value of [M/H]. $w$ stands for the relative contribution of the disc-like component with the relative contribution of the halo-like component defined as (1-$w$). $knots_w$ and $knots_\mu$ stands for the [M/H] spline knots location for the relative contribution parameter ($w$) and the velocity/circularity means and standard deviations ($\mu$, $\sigma$) respectively. The graphical representation of the models are shown in Figure \ref{fig:graphic-model}.
\end{table*}

\begin{figure*}
    \centering
    \includegraphics[width=\textwidth]{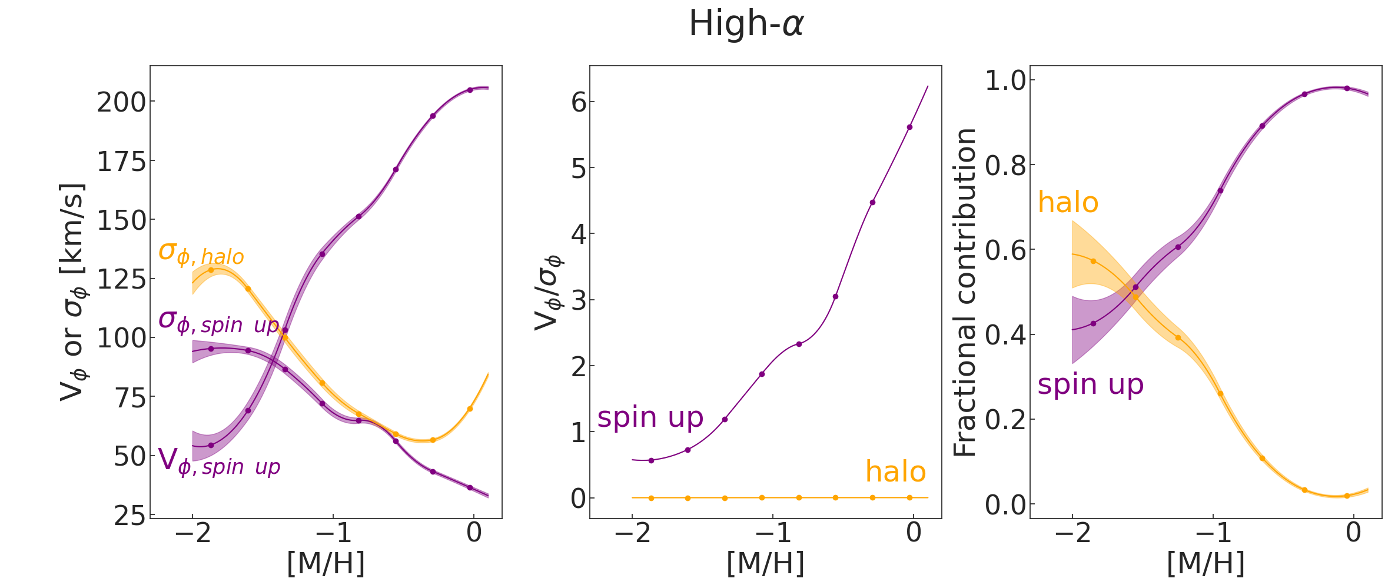}
    \caption{High-$\alpha$ model with evolving means and standard deviation for azimuthal velocity of the disc-like component. Left: Evolution of azimuthal velocity and azimuthal velocity dispersion as a function of metallicity, capturing a monotonic transition of a chaotic dispersion-dominated state (predominantly made of proto-Galactic populations) to a rotationally supported state with disc-like motion in purple;  the velocity dispersion evolution of halo is shown in orange. Centre: Evolution of azimuthal velocity over azimuthal velocity dispersion, V$_\phi$/$\sigma_\phi$ as a function of metallicity for the disc phase and halo component is shown in purple and orange, respectively. Right: Fractional contribution of disc phase (proto-Galaxy to high-$\alpha$ disc) (purple) and halo-like (orange) GMM components as a function of metallicity. In all the panels, the scatter points show the location of the knots chosen to run our model. The bands are 1$\sigma$ errors on the converged weights, velocities, and velocity dispersions. The fractions are computed for an evolving Gaussian for the disc phase and fixed Gaussian for the halo phase when the MCMC is converged.}
    \label{fig:frac-model2}
\end{figure*}

\begin{figure*}
    \centering
    \includegraphics[width=\textwidth]{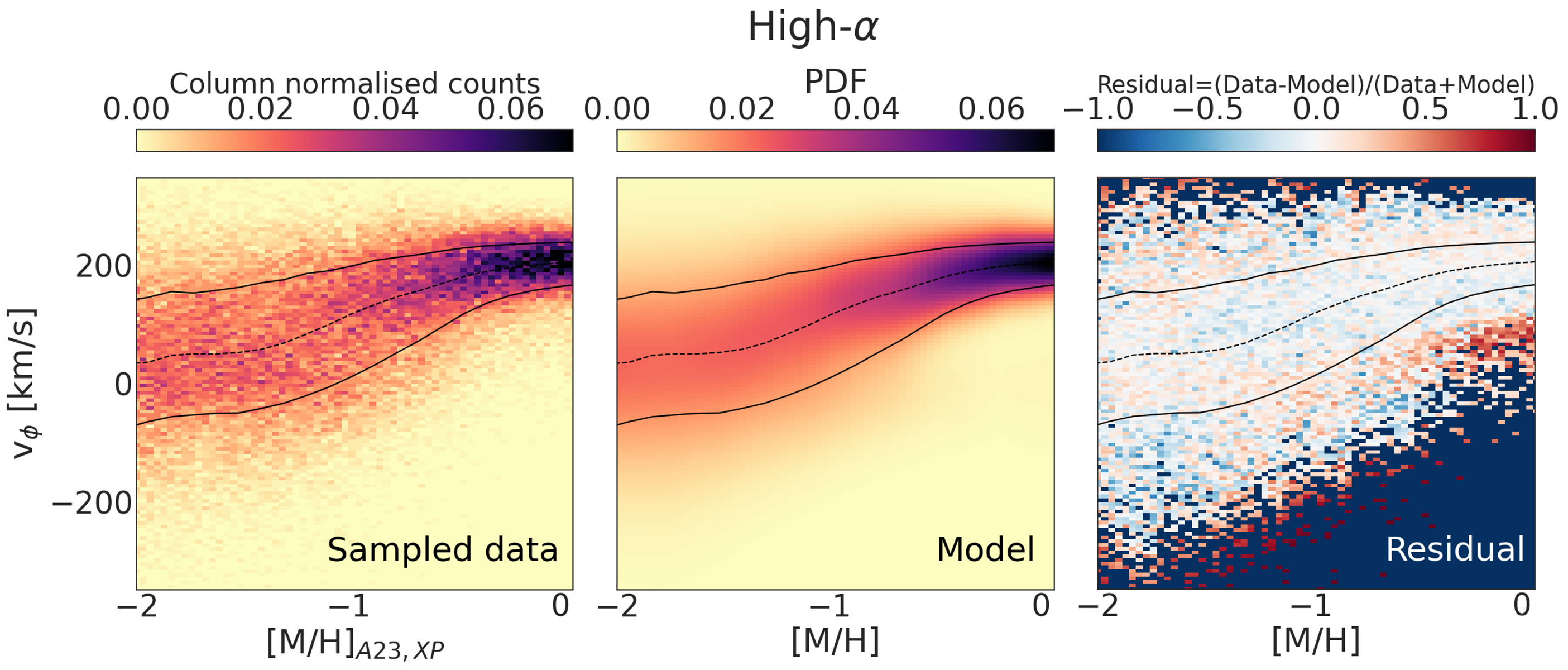}
    \caption{Distribution of sampled data column normalised by sum (left), model colour-coded by the probability density function (centre), and the normalised residual (right) in the [M/H]-v$_\phi$ plane (azimuthal velocity vs metallicity) for high-$\alpha$ evolving means model. The flat distribution of red stars at higher metallicities in the residuals show that the model is not able to fit the non-Gaussianity (long tail towards lower velocities) of the high-$\alpha$ disc. The systematic scatter in the periphery is due to random error in the data. However, for [M/H]<--0.7, the lack of strong systematic patterns, the low-amplitude, and the small scatter of the residuals in the central regions (with the major bulk of the data) verifies the validity of the model. The median, 16${th}$ and 84${th}$ percentile tracks for our data are shown in black lines to guide the eye towards the bulk of the data.}
    \label{fig:residual}
\end{figure*}

In order to quantify the spin-up of the high-$\alpha$ disc population, we modify the simple two-component model from the previous section to allow the mean velocity and velocity dispersion to vary between the spline knots across the metallicity range.
We fix the mean velocity of the halo component to be 0 km s$^{-1}$ across the metallicity knots (the value converged from the HMC run using fixed means and standard deviations). 
This is fixed in order to separate the metal-poor high-$\alpha$ in situ population (likely components of the proto-Galaxy, with a small but non-negligible prograde rotation) from the more radial halo population, as they heavily overlap.
However, we let the standard deviation of the halo-like component vary across the metallicity knots with a truncated normal prior with a mean of 100 km s$^{-1}$ and standard deviation of 50 km s$^{-1}$, with lower and upper limits restricted between 50 and 200 km s$^{-1}$.

It is important to note that the high-$\alpha$ selection is not perfect, and still captures the metal-poor end of accreted mergers (like GES).
Therefore, fitting the halo-like component is still important in our high-$\alpha$ sample.
The mean and standard deviation of the velocity of the disc-like component are set to vary with increasing metallicity. This will allow us to measure the spin-up of a proto-Galactic population to a rotation-dominated high-$\alpha$ disc. 
The metallicity range used for high-$\alpha$ stars with this model is between --2.0 and +0.1. 
For the more metal-rich bins, we undersample the data to have an equal number of stars in each bin. The number of stars is set to the number in the lowest metallicity bin. 
This downsampling reduces the sample size for the most metal-rich bins, that in turn reduces the computation time.
This also allows us to have equidistant knots, allowing the sampling of the metal-poor end as well as the metal-rich end (as opposed to the log-space knots used in the previous subsection, which sample the metal-poor end less than the metal-rich end).
We choose 8 metallicity knots (equidistant in linear space) for the relative fraction and 9 metallicity knots (equidistant in linear space) for the velocity means and standard deviations. The mean velocity, velocity dispersion, and relative weights are computed for each knot in metallicity and spline interpolated for the data in-between the knots. 

In order to measure the spin-up of the disc-like component, and to enable the information of the disc-like component to be carried across metallicity bins, we implement a Gaussian process such that every finite collection of the azimuthal velocities (measuring the mean azimuthal velocity of the disc-like component) indexed by its metallicity has a multivariate normal distribution described by a rational quadratic kernel function\footnote{This is because Gaussian processes can be seen as an infinite-dimensional generalisation of multivariate normal distributions.} 

\begin{equation}\label{eq3:cov}
    cov_{\rm spin-up}(\rm [M/H]_i,\rm [M/H]_j) = \sigma^2\left(1+\frac{([M/H]_{i}-[M/H]_{j})^2}{2\alpha\ell^2}\right)^{-\alpha},
\end{equation}

where $\sigma^2$ is the overall variance, $\ell$ is the characteristic length scale of the covariance function, that describes the range in [M/H] to which the covariance function typically holds, and $\alpha$ is the positive scale-mixture parameter of the covariance function ($\alpha$ > 0), that in simple terms, describes the curvature of the covariance function.
This function models the covariance between each pair of ([M/H]$_i$,[M/H]$_j$).
The standard deviation on the mean of the azimuthal velocity describing the disc-like component has a uniform prior between 50 and 200, the characteristic length scale has a uniform prior between 0 and 1, and the scale-mixture parameter has a uniform prior between 0 and 4. 
All these parameters are sampled with the MCMC along with the means, standard deviations, and weights of the disc-like and halo-like components across the metallicity bins.
The mean azimuthal velocity of the disc-like component has a multivariate normal distribution prior centred at 120 km s$^{-1}$ described by the rational quadratic kernel function covariance matrix shown in equation \ref{eq3:cov}.
The mean azimuthal velocity is therefore defined as the mean function together with the kernel function that define the Gaussian process distribution of the azimuthal velocity of the disc-like component with varying metallicity. 
The standard deviation is described by a half normal prior centred at 150 km s$^{-1}$.
The relative weight of the disc-like component is forced to be monotonically increasing using the \texttt{positive\_ordered\_vector} constraint on the argument, which automatically forces the halo-like component to be monotonically decreasing, removing noisy fits in the low-metallicity end. 

The NUTS sampler with a HMC inference is run on the sampled data with the model described above for 100 warm-up steps, and 50 samples to generate from the Markov chain for the high-$\alpha$ stars.
We show the graphical model representation of this model in the right panel of Figure \ref{fig:graphic-model} and a list of model parameters and their functional forms are provided in the third column of Table \ref{func-form}.
The high-$\alpha$ sample chains look much more stable and converged with this evolving velocity means and standard deviations model, they have \texttt{r\_hat} below 1.1 (between 0.98 and 1.03), and \texttt{n\_eff}, the number of effective sample is at least 30 or more for the means, standard deviations and relative weights, and about 20 for the covariance function parameters.
The converged characteristic length is 0.75 (showing larger correlation between the velocities at different metallicities), and the scale-mixture parameter is 1.95. In summary, these results suggest that this model provided posterior distributions that better describe the underlying high-$\alpha$ sample.

Figure \ref{fig:frac-model2} shows the converged parameters as a function of metallicity. As in Figure \ref{fig:frac-model}, the metallicity knots are shown as scatter points and $1\sigma$ uncertainties on the converged parameters are shown as coloured bands.
The disc-like component is shown in purple and the halo-like component is shown in orange.
The left panel of Figure \ref{fig:frac-model2} shows the evolution of the mean azimuthal velocity and azimuthal velocity dispersion for the high-$\alpha$ disc-like component. Here, the trend gradually increases from v$_{\phi}\sim50$ km s$^{-1}$ at [M/H] $\sim-2$ to v$_{\phi}\sim200$ km s$^{-1}$ at [M/H] $\sim0$. This result can be interpreted as the high-$\alpha$ disc spinning up from a proto-Galactic population to a rotation dominated disc. While previous work have alluded to this in the literature \citep{2023chandra,2023zhang}, this is the first time that the evolution of the high-$\alpha$ proto-Galaxy-to-disc population has been quantitatively measured. In more detail, our results suggest that the proto-Galactic phase (pre-disc) lasts for approximately 0.3 dex (between --2 and --1.7 dex). After this, the proto-Galaxy populations gain azimuthal velocities as metallicity increases ---in an almost linearly fashion--- until they settle into a disc at around --0.5 dex. In summary, our results show that the so-called ``spin-up'' phase of the Galaxy happens gradually across a large range of [M/H], starting from metallicities as low as [M/H] $\sim-1.7$.

Furthermore, throughout this phase, the velocity dispersion of the disc-like component decreases with metallicity, which also highlights how as the population gains v$_{\phi}$, it becomes less dispersion dominated. 
The velocity dispersion of the halo component (orange, that has fixed mean of v$_{\phi}=0$ km s$^{-1}$) is also decreasing with increasing metallicities. This could be due to different substructures dominating different metallicities. 
For example, we know that the GES merger has a lower velocity dispersion (of about 50 km s$^{-1}$) compared to the rest of the halo resting at about 100 km s$^{-1}$ velocity dispersion (see Figure~\ref{fig:median-vphi-feh}). 
The rise in standard deviation at higher metallicities (when the high-$\alpha$ disc is in place) is not physical, but is rather caused by the model trying to make a broad halo component to fit the asymmetric tail of the high-$\alpha$ disc. This is one of the disadvantages of using a Gaussian distribution.

The middle panel of Figure \ref{fig:frac-model2} shows the ratio of v$_\phi$ to $\sigma_\phi$, measuring how rotationally supported or 'discy' the stars are. 
This ratio is basically zero (extremely pressure-supported) for the halo-like component, as we set the mean of the halo-like component to be close to 0 km s$^{-1}$.
However, the disc-like component has a clear rise in v$_{\phi}/\sigma_{\phi}$, reaching up to a ratio of 6 at solar metallicities. 
Moreover, the right panel of Figure \ref{fig:frac-model2} shows the relative contribution of halo-like (orange) and disc-like (purple) components at different metallicities. 
The accreted halo contribution decrease quickly with increasing metallicities.
It dominates the distribution only at the lowest metallicity bins, below [M/H]$\lesssim-1.7$.
In the very metal-poor end ([M/H]<--2) of our high-$\alpha$ selection, the halo-like to disc-like component (i.e. halo-like to proto-Galaxy-like) ratio is 40\%:60\%. 
It is worth noting that we do not call this component "disc", but simply refer to this component as "disc-like" for the modelling purpose.
This component captures the proto-Galaxy to spin-up phase to high-$\alpha$ disc.
The disc-like component's fractional contribution increases slowly from [M/H]$\sim-2$, with an approximately constant gradient, up to [M/H]$\sim$--0.5. Upon reaching this point, the disc-like component (i.e. high-$\alpha$ disc) dominates the sample.
All these results are highly in favour of a gradual spin-up for the disc-like component with increasing metallicities. In terms of the evolution of the Galaxy, our results highly favour the scenario where a proto-Galaxy population with low (but non-negligible) v$_{\phi}$ profile spins up into a fully rotation-dominated high-$\alpha$ disc.

In Figure \ref{fig:residual}, we show the [M/H]-v$_\phi$ plane for high-$\alpha$ sample, with the sampled data in 2D column-normalised histogram on the left, the fitted model normalised with the integral under the curve equal to 1 (similar to column normalisation by sum) in the middle, and normalised residual (i.e. data -- model) on the right.
All the panels have the running 16${th}$, 50${th}$, and 84${th}$ percentile tracks from Figure \ref{fig:median-vphi-feh} shown as black lines.
We can clearly see that the sampled data follows the median tracks very well.
We construct the model using the splines on the mean velocities, velocity dispersions, and fractional contributions for the two-component fits.
The normalised residual is constructed by subtracting the probability density function of the sampled data with the model's probability density function (PDF) in each cell (with bin sizes of 6.25 km s$^{-1}$ in v$_\phi$ and 0.04 in [M/H]).
The residuals are scaled by the sum of sampled data and the model's PDF. 
This way, the residual only goes from -1 to 1 in value.
If the residual is less than 0 the model predicts more stars than the data shows, and if the residual is greater than 0 the data has more stars in that region than the model predicts.

When inspecting Figure \ref{fig:residual}, the first thing one catches by eye in the residuals (right panel) is the horizontal patch of red stars at higher metallicities (between --0.5 < [M/H] < 0). Here, the model predicts fewer stars than what is present in reality. We conjecture this is because the v$_\phi$ distribution of high-$\alpha$ disc stars is asymmetric probably caused by the mechanism of asymmetric drift and is strongly non-Gaussian with a long tail towards lower v$_\phi$ for this metallicity bin. Thus, the model is not able to capture well these stars.
This asymmetric drift is stronger in high-$\alpha$ disc than the low-$\alpha$ disc. However, it is present in both populations \citep{2020Anguiano}. The same effect can also be seen in the model with frozen means and standard deviation for azimuthal velocity in the low-$\alpha$ sample.
Due to this asymmetric tail towards lower velocities, we find that the model is underfitting the data by 22-26\% at higher metallicities ([M/H]>-0.5), while the under/overfitting is as low as 2-5\% at lower metallicities.

The evolving mean high-$\alpha$ model presented here models well the bulk of the stars (also represented by the percentile tracks). However, we note that it struggles to fit the stars in the periphery of the v$_\phi$ distribution (edge of the grid values in [M/H]-v$_\phi$ plane), mostly due to Poisson error. 
When compared to the residuals for the model with frozen means and standard deviation for azimuthal velocity in the high-$\alpha$ sample, the evolving means model has much smaller systematic effects. This is due to the frozen means model's underlying assumption of frozen mean velocity and velocity dispersion. 
Therefore, this improved model with evolving means is a much better representation of the high-$\alpha$ sample. 
Furthermore, it is important to note that in the residuals, the model is fully represented by a gradual (almost linear) rise in azimuthal velocity over the entire range of metallicities. If the data/model were better represented by a rapid and more exponential growth of v$_{\phi}$ with respect to a narrow range of metallicity --as reported in the literature--, we would find large systematic effects in our residuals (that are not seen). The absence of such systematic effects is more supporting evidence that the metal-poor high-$\alpha$ (or proto-Galaxy population) gradually spins-up to a rotation-supported high-$\alpha$ disc over a wide range of metallicities.

To our knowledge, this model is a first attempt at a simplified, physically motivated, and easily interpretable representation of the azimuthal velocity versus metallicity plane in the high-$\alpha$ regime, capturing the evolution of the first 5-6 Gyr of the proto-Milky Way populations to the high-$\alpha$ disc. 
However, given the consequences of the simplified Gaussian distribution assumption, this model is a more qualititative representation in the metal-rich end ([M/H]>--0.5). 

\section{Evolution of orbital circularity with [M/H]}\label{4.1}

\begin{figure*}[h!!!!]
    \centering
    \includegraphics[width=\textwidth]{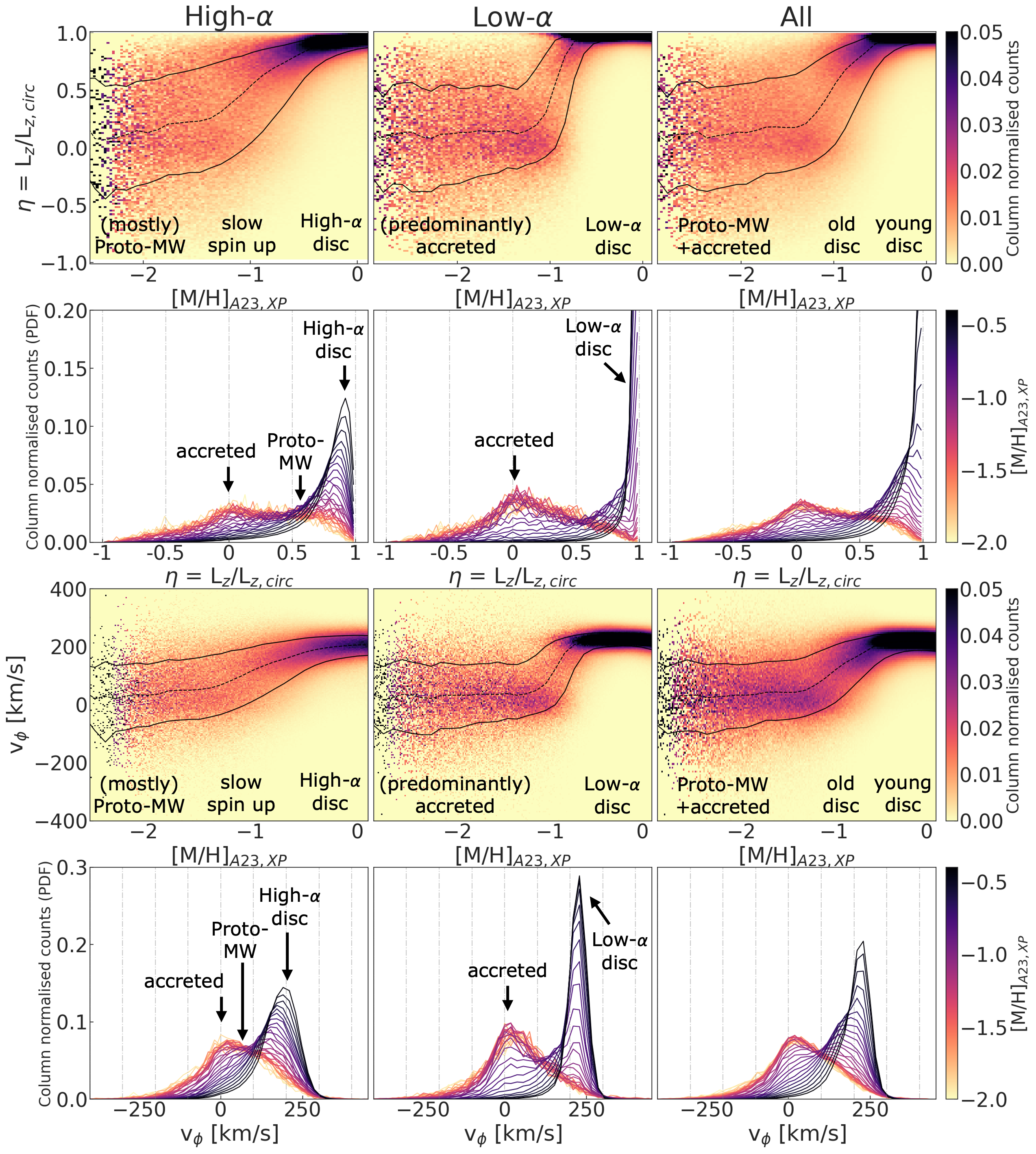}
    \caption{Top: Column-normalised (by sum) 2D histogram of stars in the [M/H]-$\eta$ plane (orbital circularity vs metallicity, top) and [M/H]-v$_\phi$ plane (azimuthal velocity vs metallicity, middle bottom) for all the stars (right), high-$\alpha$ selection (left), and low-$\alpha$ selection (midde). The running median track is shown as dashed black line and the 16${th}$ and 84${th}$ percentile tracks are shown as black lines in all panels. The running median tracks for the \textit{all stars} and low-$\alpha$ panels look more like a step-function, while the high-$\alpha$ tracks are shallower, supporting the more gradual spin-up phase with respect to metallicities. (middle top) Probability density functions (PDF) of $\eta$ in bins of [M/H] for all the stars (right), high-$\alpha$ selection (left), and low-$\alpha$ selection (middle). (bottom) Probability density functions (PDF) of v$_\phi$ in bins of [M/H] for all the stars (right), high-$\alpha$ selection (left), and low-$\alpha$ selection (middle). We note the second peak with low net spin in high-$\alpha$ panel slowly gains rotation with increasing [M/H] and spins-up into the old high-$\alpha$ disc in both middle top and bottom panels.}
    \label{fig:circ-main}
\end{figure*}

The orbital circularity metric, $\eta$, defined with respect to the present-day disc alignment, ranging from --1 (perfectly retrograde, in-plane) to 1 (perfectly prograde, in-plane), interprets a star's orbital properties within the Galaxy. 
Intermediate values represent elliptical orbits (closer to 0, isotropic, radial).
This quantitative measure of orbital shape is very useful in understanding the dynamical processes that shaped the formation of the high-$\alpha$ disc. 
We compute orbital circularities for our sample of stars using the \texttt{gala} software package \citep{2017pricewhelan}. We integrate the orbits of stars within a realistic Milky Way model for the Galactic potential, \texttt{MilkyWayPotential2022} \citep{2022pricewhelan}, which incorporates four crucial components: a dense central core, a surrounding bulge of stars, a flattened disc of stars and gas, and a vast, spherical dark matter halo. 
This model is also calibrated to match observations of the Milky Way's rotation curve by \citet{2019eilers} and a compilation of Milky Way's total mass enclosed \citep[see][]{2022hunt}. 
Using the adopted gravitational potential, we calculate the vertical angular momentum (L$_{z, circ}$) and total energy (E) for a perfectly circular orbit. 
By interpolating the resulting L$_{z,circ}$(E) curve for each observed star's total energy, we determine the orbital circularity as $\eta$ = L$_z$/L$_{z,circ}$.

The evolution of orbital circularity with respect to increasing metallicities is shown as 2D column normalised (by sum) histograms in the top panels of Figure \ref{fig:circ-main} for high-$\alpha$, low-$\alpha$, and \textit{all stars}.
All the 2D histograms have 16${th}$, 50${th}$, and 84${th}$ percentile tracks of $\eta$ versus [M/H] overlaid.
To compare the orbital circularity with the azimuthal velocities used in this work, the 1D histograms of azimuthal velocities in bins of metallicities are shown in the bottom panels of Figure \ref{fig:circ-main} for the high-$\alpha$, low-$\alpha$, and all samples.
We also show the 1D histograms of orbital circularity, $\eta$, in bins of metallicities in the second row of Figure \ref{fig:circ-main}.
The use of orbital circularity is very important because, unlike v$_\phi$, orbital circularity is position independent (especially at larger distances away from the solar neighbourhood).
From the high-$\alpha$ panel of Figure \ref{fig:circ-main}, we can confirm the gradual evolution of circularity from a non-zero median circularity ($\eta\sim0.1$) metal-poor (halo-like) population to a rotation-dominated ($\eta\sim0.9$) disc-like component over a broad range of metallicities, ranging between --2.5 < [M/H] < --0.7 (see the median tracks overlaid), if they are composed of a single stellar population.
This is in favour of the gradual spin-up phase of the Milky Way's high-$\alpha$ disc over increasing metallicities.
This can also be seen in the 1D histograms of v$_\phi$ (fourth row) and $\eta$ (second row), with the lighter colours (lower metallicities) having a bimodal distribution; this bimodal feature is likely attributed to the superposition of a halo-like population with no rotation and a proto-Galaxy-like population with small rotation \citep{2024horta}.
The bimodal distribution at lower metallicities is strikingly clear for orbital circularities as annotated in the second row of Figure \ref{fig:circ-main}.
Furthermore, the low-$\alpha$ panel of Figure \ref{fig:circ-main} reveals that the low-$\alpha$ stars are made of two (almost) disjoint populations: an accreted halo and low-$\alpha$ (thin) disc. The halo dominates the lower metallicity bins, whilst the low- $\alpha$ disc dominates the higher metallicities, as expected.
This can be seen also in the median tracks (top panel of Figure \ref{fig:circ-main}), which delineate a trajectory similar to a step function.
The low-$\alpha$ 1D histograms of v$_\phi$ also show that the accreted halo (lighter colours, lower metallicities) dominates at v$_{\phi}\sim0$ km s$^{-1}$ from [M/H] $\sim-2$ to [M/H] $\sim-1$, without a decrease in the relative weight (i.e. the peak in the distribution stays relatively the same. After [M/H] $\sim-1$, the v$_\phi$ distribution rapidly changes into a highly rotating, disc dominated, population, that spans over 0.7 dex in metallicity (darker colours, higher metallicities).

In the \textit{all stars} bottom panel of Figure \ref{fig:circ-main}, we see a combination of the high-$\alpha$ and low-$\alpha$ components: accreted halo+proto-Galactic populations dominating the lower metallicities, high-$\alpha$ disc populations emerging from --1 < [M/H] < --0.7, and low-$\alpha$ disc populations taking over the distribution at metallicities higher than [M/H] $\sim-0.6$.
The same picture can be deduced from the 1D histograms of v$_\phi$, that trace the same distribution as the circularity. 
Moreover, the median tracks on the 2D histograms show a rapid rise in circularity at around [M/H] $\sim-1.0$, which is driven by a transition from the accreted (mostly GES) debris to the high-/low-$\alpha$ discs, similar to what is seen in the low-$\alpha$ sample. However, in the \textit{all stars} panel, the jump is not as sharp due to the presence of high-$\alpha$ disc populations as well.
The conclusions are therefore consistent with the use of orbital circularity or azimuthal velocity.
Given the position dependence on the azimuthal velocity, the use of orbital circularity brings more confidence that the gradual rise in rotational support over increasing metallicities for the high-$\alpha$ population is bonafide. 

This result is important. If one does not account for the $\alpha$-separation like done in this work, a conclusion of the disc spin up in the [M/H]-v$_{\phi}$ diagram could be interpreted as rapid, which we find is not the case (see Figure \ref{fig:circ-main}). The rapid transition from radial (v$_{\phi}\sim0$ km s$^{-1}$) orbits to circular ones (v$_{\phi}\sim200$ km s$^{-1}$) is caused by the presence of accreted populations and the low-$\alpha$ disc, and is only seen when examining low-$\alpha$ populations. On the contrary, when looking solely at high-$\alpha$ populations --which should trace directly the spin up of the old proto-Galaxy to the  high-$\alpha$ disc--, it is immediately clear that the relation in this diagram is much more gradual with respect to increasing metallicities. 

Previous work has attempted to look at the transition between hot/radial orbits and cold/circular ones using this $\alpha$-selection \citep{2023chandra}. Thus, it is important that we compare our results to theirs. We argue that the reason these [M/H]-$\eta$ 2D column-normalised histograms look strikingly different (especially for the high-$\alpha$ samples) from the study by \citet{2023chandra} is because of the way the 2D histograms are plotted (both studies use the same data and similar $\alpha$-separation curves). \citet{2023chandra} column-normalises their histograms by amplitude (tracing the mode of the distribution), while we column-normalise by their sum (tracing the underlying PDF). Column-normalisation by amplitude traces the mode of the distribution, which makes the whole 2D histograms more noisy (as the distribution is scaled by a factor of the standard deviation of the curve). Tracing the mode also means that in a bimodal distribution across a large range of metallicities (like for the high-$\alpha$ sample), the mode switches between one and the other rapidly within a small bin size in metallicity. This could lead to the data manifesting a sharp increase from $\eta\sim0$ to $\eta\sim1$, as seen in \citet{2023chandra}, when in reality the data shows a more gradual increase in circularity with respect to metallicities (this work). Therefore, it is important to know how different normalisation methods can give rise to different interpretations of the underlying data and choose the normalisation method that best represents the science question that needs solving. In our case, as we aim to understand how the low-[M/H] regime transitions into the high-[M/H] regime for both high-/low-$\alpha$ populations, we choose to plot the sum column-normalised distribution. Furthermore, the differences with \citet{2023chandra}'s results also arise from a small difference in the $\alpha$-separation. 
These normalisation and $\alpha$-selection differences and their implications are discussed in detail in Appendix \ref{B}.


\begin{figure}
    \centering
    \includegraphics[width=\columnwidth]{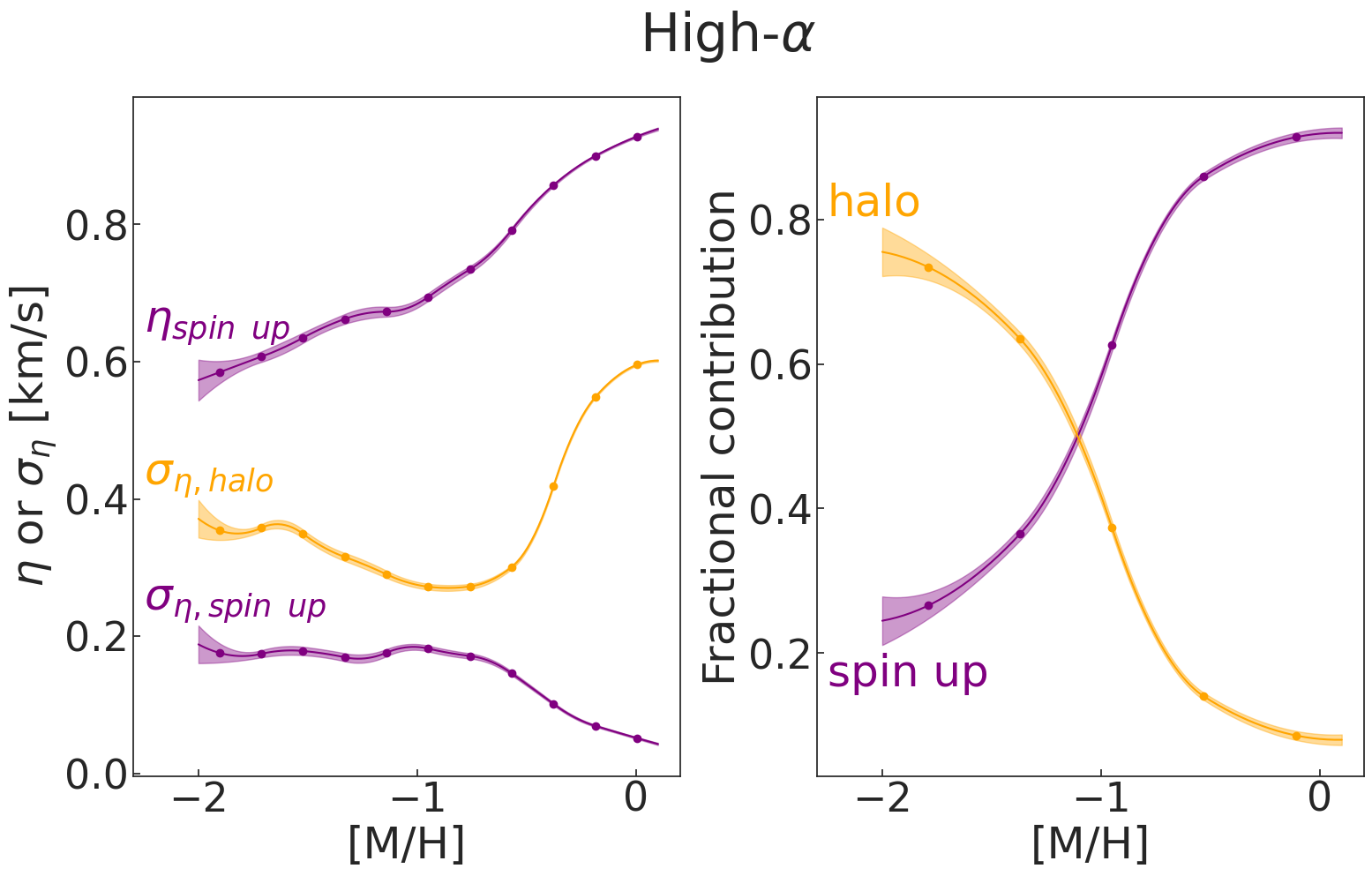}
    \caption{High-$\alpha$ model with evolving means and standard deviation for orbital circularity of the disc-like component. Left: Evolution of orbital circularity, and orbital circularity dispersion as a function of metallicity, capturing a gradual transition of a chaotic state (predominantly made of proto-Galactic populations) to a rotationally supported disc-like motion in purple and the orbital circularity dispersion evolution of halo is shown in orange. Right: Fractional contribution of disc phase (proto-Galaxy to high-$\alpha$ disc) (purple) and halo-like (orange) GMM components as a function of metallicity. In all the panels, the scatter points show the location of the knots chosen to run our model. The bands are 1$\sigma$ uncertainties on the converged weights, circularity, and circularity dispersions. The fractions are computed for an evolving Gaussian for the disc phase and fixed Gaussian for the halo phase when the MCMC is converged.}
    \label{fig:circ-model}
\end{figure}

Even though orbital circularity depends on the adopted Galactic potential, it is position independent as opposed to azimuthal velocity.
This makes it valuable to model the circularity evolution across metallicities.
Due to the non-Gaussian nature of orbital circulairity (as it abruptly cuts at -1 and 1 for a perfectly circular retrograde and prograde orbits respectively), the models described earlier in this work using azimuthal velocities cannot be directly applied to orbital circularities.
Therefore, the underlying Gaussians are modelled as folded normal distributions (folded at -1 and 1).
The modelling is only performed on high-$\alpha$ stars because they trace the transition of the proto-Galaxy to rotation-supported high-$\alpha$ disc more clearly as seen in Figure \ref{fig:circ-main}, while the low-$\alpha$ stars can be simply described by two separate components similar to the [M/H]-v$_\phi$ plane.

The halo-like component has a fixed mean of 0 while varying standard deviation  with a truncated normal prior with a mean of 0.4 and standard deviation of 0.2, with lower/upper limits restricted between 0.2 and 0.7 across the metallicity knots. 
We use the same metallicity range as in subsection \ref{3.3.1}, with 12 and 6 metallicity knots (equidistant in linear space and downsampled between each bins to have the same number of stars) for relative fractions and orbital circularity means and standard deviation.
The definition of a covaraince matrix to enable the information between each components across the metallicity is unchanged in this circularity model, with the standard deviation on the mean of the circularity having a uniform prior between 0.1 and 0.9.
The mean orbital circularity of the disc-like component is described by a multivariate normal distribution centred at 0.3 and the standard deviation is described by a half normal prior centred at 0.3.
The final distribution of both the components are converted to a folded normal distribution to account for the folding at -1 and 1.  

The model is run with 250 warm-up steps and 50 sample chains with the chains converged and \texttt{r\_hat} less than 1.1 and effective samples above 30. 
We show the graphical model representation of this model in the right panel of Figure \ref{fig:graphic-model} and a list of model parameters and their functional forms are provided in the fourth column of Table \ref{func-form}.
Figure \ref{fig:circ-model} shows the converged parameters as a function of metallicity after the spline interpolation between the metallicity knots.
We see that the orbital circularity of the disc-like component is steadily increasing with increasing metallicities over a large range of metallicities, [M/H] $\sim$ --1.7 to --1 and is still steadily increasing up to [M/H]$\sim$--0.5.
However, the interpretation of metal-rich stars ([M/H]>--1.0) is trickier given that the long asymmteric tail of the disc-like population is poorly fit due to the assumption of the underlying distribution as a simple (folded) Gaussian distribution.
For the metal-poor stars, we can more confidently say that the spin up phase is more extended over a large range of metallicities from [M/H] $\sim$ --1.7 to --1. 
The relative fractional contribution from the spin up (disc-like) and halo-like population also supports the slower evolution of circularity with respect to metallicities, as opposed to the rapid spin up shown in the literature \citep[see  e.g.][]{2023chandra,2024kurbatov}.
The major difference between the azimuthal velocity and orbital circularity evolution from the modelling perspective is (i) the metallicity at which the halo and spin up component crossover in the relative contribution, which is much more metal-rich in circularity ([M/H] $\sim$ --1.2) compared to velocity ([M/H] $\sim$ --1.7), and (ii) the mean orbital circularity is more circular ($\eta$ $\sim$ 0.57) at lower metallicities ([M/H] < --1.5, also seen in the 1D histogram in Figure \ref{fig:circ-main}) than previously reported and more circular than what mean azimuthal velocities show at the same metallicities (v$_\phi$ $\sim$ 50 km s$^{-1}$).
Both of these differences can be explained due to the fact that circularity has less position dependence than velocities (given that our sample has stars outside the solar neighbourhood; 49\% of our stars are at heliocentric distances larger than 3 kpc).
This is because the mean azimuthal velocity reduces close to the inner Galaxy when compared to solar neighbourhood, making the mean azimuthal velocity smaller in value.
This affects the metal-poor stars more, as metal-poor stars are centrally concentrated \citep{2022rix}, especially in the high-$\alpha$ sample.
The latter difference could also arise from the fact that in the literature, the metal-poor high-$\alpha$ stars are not modelled with both an accreted and in situ population, whereas we can clearly see a bimodal distribution in the 1D histograms in Figure \ref{fig:circ-main}, justifying our choice of modelling the leftover accreted halo stars in the high-$\alpha$ regime along with the proto-Milky Way-like population.
Therefore, the evolution of orbital circularity over increasing metallicities shows that the transition from a slowly rotating population to high-$\alpha$ disc is more gradual and not as rapid at [M/H] $\sim$ --1.
However, an important caveat to mention is that the orbital circularity represents how circular the orbit of a star is compared to the present-day disc orientation, which does not necessarily trace the circularity at formation.

\section{Discussion}\label{4}

In this section, we discuss the summary of our results and the implications of the gradual spin-up phase across a large range in metallicities. 
We furthermore perform a simple GMM decomposition in bins of metallicities, to support the gradual spin-up phase scenario.
Finally, we present the limitations and future scope of this work.

\subsection{Summary of results}

In this work, we have set out to model the azimuthal velocity and orbital circularity evolution of high-/low-$\alpha$ stars across metallicity space using \textsl{Gaia} XP element abundances, DR3 astrometry, and RVS radial velocites, to understand the transition phase of the proto-Galactic population to the high-$\alpha$ disc. 
At first, we model the conditional distribution P(v$_\phi \mid $[M/H]) using a two-component (disc-like and halo-like) GMM with frozen means and standard deviations for the azimuthal velocities for the high- and low-$\alpha$ stars.
We find the inferred posterior matches better with the data for low-$\alpha$ stars than high-$\alpha$ stars.
The fractional contribution from disc-like and halo-like components look like a step function at [M/H]$\sim$--1 for low-$\alpha$ stars, while the high-$\alpha$ stars have a relatively gradual rise over increasing metallicities in the fractional contribution (Figure \ref{fig:frac-model}).
Secondly, given that the high-$\alpha$ stars are not modelled well by the frozen means model, we model the conditional distribution P(v$_\phi \mid $[M/H]) using a two-component (disc-like and halo-like) GMM with evolving means and standard deviations for the azimuthal velocities for the high-$\alpha$ stars.
From this exercise, we see that both the mean azimuthal velocity and fractional contribution of the disc-like component is gradually increasing starting from a v$_\phi$ $\sim$ 50 km s$^{-1}$ over increasing metallicities (--1.7<[M/H]<--1.0).
Thirdly, we perform the same exercise for high-$\alpha$ stars in orbital circularity versus metallicity plane, [M/H]-$\eta$, given that the orbital circularity is position independent, as opposed to azimuthal velocity.
From this exercise, we also see a gradual rise in oribital circularity starting from an $\eta\sim$ 0.57 over increasing metallicities for the disc-like component.

Using different flavours of these mixture models, we have provided several lines of evidence that the metal-poor high-$\alpha$ disc increases its average azimuthal velocity, orbital circularity and rotational support gradually and monotonically across a wide range of [M/H], spanning approximately $-1.7 <$ [M/H] $< -1$. 
These data favour the scenario of a gradual spin-up of the metal-poor high-$\alpha$ disc (likely the proto-Galaxy) to a rotationally supported high-$\alpha$ disc at higher [M/H]. 
On the contrary, due to the superposition of the GES debris and the low-$\alpha$ disc in the low-$\alpha$ sample, the transition from metal-poor (halo) populations to metal-rich (disc) ones is much sharper, appearing almost like a step-function at [M/H]$\sim-1$. 
Due to the GES debris dominating the metal-poor sample for \textit{all stars} in our dataset, this yields a similar profile when inspecting the \textsl{Gaia} XP sample without any $\alpha$ selection. Thus, our results highlight the importance of the [$\alpha$/M] selection for studying the azimuthal velocity evolution of the old Milky Way disc and to avoid the GES debris.
This also suggests that the proto-Galactic debris gained rotation gradually over increasing metallicities, eventually settling into a high-$\alpha$ disc and the disc formation was not as rapid or dramatic across a narrow range of metallicities as previously reported.

\subsection{On the spin up of the Milky Way disc}\label{4.3}

A comprehensive understanding of disc evolution necessitates accounting for the complex interplay between stellar metallicity, $\alpha$-element enhancement, age, and spatial distribution. Foundational models such as the monolithic collapse scenario proposed by \citet{1962eggen} described the Milky Way forming rapidly from a single collapsing gas cloud. While this framework was influential, it cannot fully account for the detailed chemical and structural features observed today--particularly the clear dichotomy between the thin and thick disc components and the chemical bimodality in the $[\alpha/Fe]$--[Fe/H] plane. To address these limitations, models like the two-infall scenario \citep{1997chiappini} introduced sequential gas accretion episodes, offering a more dynamic and layered view of disc formation. As highlighted in recent studies \citep[e.g.][]{2019spitoni}, combining chemical abundances with kinematic, spatial, and age information has been essential for revealing the Milky Way's more intricate evolutionary pathways. This broader context informs our interpretation of disc spin-up mechanisms and their connection to the Galaxy’s formation history.

At the earliest cosmic times, the systems that seeded the formation of the Milky Way are conjectured to be manifested as a chaotic, unstructured, ``proto-Galactic'' population \citep[e.g.][]{Mowla2024}, shaped by continual merging of building blocks that gave birth to the Galaxy's ancient stars  \citep{2010tumlinson,2018elbadry,2024horta,2024semenov,2024xiang}. Evidence of stellar populations associated with the proto-Milky Way have been proposed in the literature \citep{2019lucey, 2020arentsen,2020reggiani, 2021horta, 2022belokurov, 2024ardernarentsen}, with a clear picture now emerging thanks to the vast \textsl{Gaia} data \citep{2022rix}. Expectations from cosmological simulations corroborate these observational findings \citep[e.g.][]{2024horta, 2024mccluskey, 2024semenov}; these suggest that at present day, the remnants of proto-Milky Way fragments should present weak but systematic net rotation with respect to. the Galactic disc (v$_{\phi}\sim50$ km s$^{-1}$), should primarily be concentrated towards the innermost Galactic regions ($r\lesssim10$ kpc), and should host stars with high [$\alpha$/Fe] abundance patterns, similar to the old Milky Way disc. 

Ascertaining the transition from a dispersion dominated population to a rotationally supported one is key to unravel the genesis of the Galactic disc. 
$\alpha$-enhanced stars trace an early phase of the Galaxy's chemical evolution, having formed before the onset of significant iron enrichment from Type Ia supernovae-typically within the first $\sim$1 Gyr \citep{1979tinsley,1986matteucci,1996raiteri}. Their elevated [$\alpha$/Fe] ratios reflect nucleosynthetic contributions from short-lived massive stars via core-collapse supernovae, prior to the delayed iron input from Type Ia events. This short enrichment timescale implies that the observed changes in angular momentum or spin among these populations reflect relatively rapid dynamical processes in the high-$\alpha$ disc, rather than a gradual evolution in Galactic timescales \citep{2013haywood,2013minchev,2019mackereth}. However, such short dynamical timescales can be manifested across a wide range of metallicities. As such, tracking the kinematic evolution of these in situ populations provides a direct view of the internal processes that shaped the early high-$\alpha$ disc component.
Published age determinations for stars in the high-[$\alpha$/Fe] disc population consistently indicate that these stars are predominantly old, typically older than $\sim$10 Gyr \citep{2016martig,2018silvaaguirre,2023Imig}. This narrow and ancient age distribution places strong constraints on the timescales for both chemical enrichment and dynamical evolution within this population. In particular, it implies that the increase in spin observed among the high-$\alpha$ stars must have occurred rapidly—within the first 1–2 Gyr of Galactic evolution—consistent with enrichment dominated by core-collapse supernovae and early dynamical settling of the high-$\alpha$ disc.
Due to the inability to currently measure stellar ages precisely for metal-poor stars, metallicity ([M/H]), a much more reliable stellar parameter estimate to determine, is often used instead. While [M/H] is not a direct tracer of age, the reason why [M/H] is used is because, under the assumption that a stellar population chemically evolves in a progressive manner, stars that form later are more enriched in metals than their previous generations; in other words, old populations tend to be metal-poor while younger populations tend to be metal-rich. Thus, [M/H] has been shown to be a useful metric to study the evolution of kinematic samples \citep[e.g.][]{2022belokurov,2022xiang,2024xiang}. This is especially the case in the current era of large-scale stellar surveys. For example, the vast \textsl{Gaia} dataset of [M/H] \citep{2023andrae} and now also [$\alpha$/M] \citep{2024li} for over two million stars has given an unprecedented (qualitative) view on the transition from the proto-Milky Way to the high-$\alpha$ disc \citep[e.g.][]{2023chandra, 2023zhang}. However, most of these studies either: ($i$) examine the profile of azimuthal velocity, v$_{\phi}$, and/or orbital circularity, $\eta$, with [M/H] agglomerating stars with different [$\alpha$/M]. This is known to be an issue due to the inclusion of stellar halo populations in the sample, that have different [M/H] and kinematic distributions; ($ii$) model the transition of dispersion dominated to rotatinally dominated populations in the [M/H]-velocities plane using disconnected GMMs (see section~\ref{4.2}), making it difficult to link components across [M/H]; ($iii)$ do not provide a quantitative measure of how the Milky Way transitioned from being dispersion dominated to rotationally dominated. 

The correlation between azimuthal velocity and metallicity in the thick disc has been established in several earlier studies. Using kinematically selected samples at high Galactic latitudes, \citet{2010spagna} first identified a steady increase in rotation velocity with [Fe/H] among thick disc stars. This trend was later confirmed for chemically defined high-[$\alpha$/Fe] populations by \citet{2017kordopatis}, who found a comparable v$_\phi$–[Fe/H] gradient. Although these earlier analyses were constrained by smaller sample sizes and narrower metallicity ranges, they provided foundational evidence of a kinematic–chemical link within the thick disc. In addition, the presence of a metal-poor thick disc ([Fe/H] $\leq$ –1.0 dex) is now well established \citep[e.g.][]{2011ruchti,2011kordopatis,2017bonaca,2019dimatteo}, extending the high-$\alpha$ population to lower metallicities than those accessible in the last decades.
\citet{2022belokurov} used high-resolution spectroscopic data from the \textsl{APOGEE} survey to study the evolution of stars around the solar neighborhood they deem in situ in the metallicity-v$_{tan}$ plane; these authors found that the Milky Way ``spun up'' rapidly between $-1.3<\mathrm{[Fe/H]}<-0.9$ (see also \citet{2023zhang} for a similar result using \textsl{Gaia} XP metallicities). Along those lines, \citet{2023chandra} used the available [$\alpha$/Fe] for the \textsl{Gaia} XP sample to study the profile of [M/H]-$\eta$, finding that the high-$\alpha$ disc emerges rapidly at $\mathrm{[M/H]}\sim -1$. 

Our analysis builds on this previous works by tracing these trends across a broader metallicity baseline and with more robust statistics, enabled by the larger, well-characterised stellar samples now available with \textit{Gaia} DR3.
In this work, we have found that when performing a similar [$\alpha$/M] cut to \citet{2023chandra}, our transition from proto-Milky Way populations to the high-$\alpha$ disc (Figure~\ref{fig:circ-main}) shows a much more gradual increase in circularity with respect to metallicities when compared to these previous works. In  more detail, both when examining and modelling the [M/H]-v$_{\phi}$ and [M/H]-$\eta$ planes, we have found that the spin up of the old Milky Way disc occurs across a much wider range in [M/H], namely between $-1.7 < \mathrm{[M/H]<-1}$. The reason for the discrepancy between our results and previous work is because: ($i$) \cite{2022belokurov} do not go down to lower metallicities ([M/H]<--1.5) and have lower number statistics compared to our work. The 1D histograms of azimuthal velocities in bins of [M/H] in \citet{2022belokurov} already shows preliminary hints of a gradual spin up (see their Figure 5); ($ii$) \cite{2023zhang} perform a traditional GMM without any $\alpha$-selection, resulting in the GES debris driving the rapid spin up inference at a narrow range in metallicities; ($iii$) \citet{2023chandra} shows a qualitative view of [M/H]-$\eta$ plane, while their high-$\alpha$ sample is contaminated by GES debris more than ours, and their column normalisation is performed by amplitude instead of sum. 
Therefore, the results presented in this work, in comparison with the recent literature, shows that the proto-Galaxy to high-$\alpha$ disc transition has likely been gradual over increasing metallicities and not as dramatic over a narrow range of metallicities as previously reported. 
While direct comparisons with high-redshift measurements of V/$\sigma$ \citep[e.g.][]{2015wisnioski} are limited by differences in methodology and redshift coverage, our results (V/$\sigma$ $\approx$ 6 and 10 for high- and low-$\alpha$ disc at [M/H] = 0 respectively) offer a $z$ = 0 anchor point for the rotational support of chemically distinct disc populations in the Milky Way.

We finish this section by raising a speculative, yet interesting point. In Figure \ref{fig:frac-model2} and Figure \ref{fig:circ-model}, we find that the profile corresponding to the ``spin up'' displays a bump in the profile at [M/H]$\approx-0.8$. We have checked that this bump is not artificially caused by our fitting methods. Interestingly, the location of this bump in [M/H] coincides with the end of the low-$\alpha$ accreted sample (primarily the GES merger debris). It is postulated that this population is the debris from a major merger in the Milky Way that dominates the local stellar halo ($6 \lesssim r \lesssim 30$ kpc: \citealp[e.g.][]{2018belokurov,2018helmi,2018koppelman}). This merger possibly catalyzed the formation of the in situ halo, or "hot high-$\alpha$ disc" comprising stars on halo-like orbits with chemistry similar to the high-$\alpha$ disc \citep{2019dimatteo,2020bonaca,2020belokurov}. Thus, we speculate that this bump could be a sign of the impact of the GES with the old high-$\alpha$ Milky Way thick disc.

\subsection{Gaussian mixture model on 3D velocities in bins of metallicities}\label{4.2}

\begin{figure}
    \centering
    \includegraphics[width=\columnwidth]{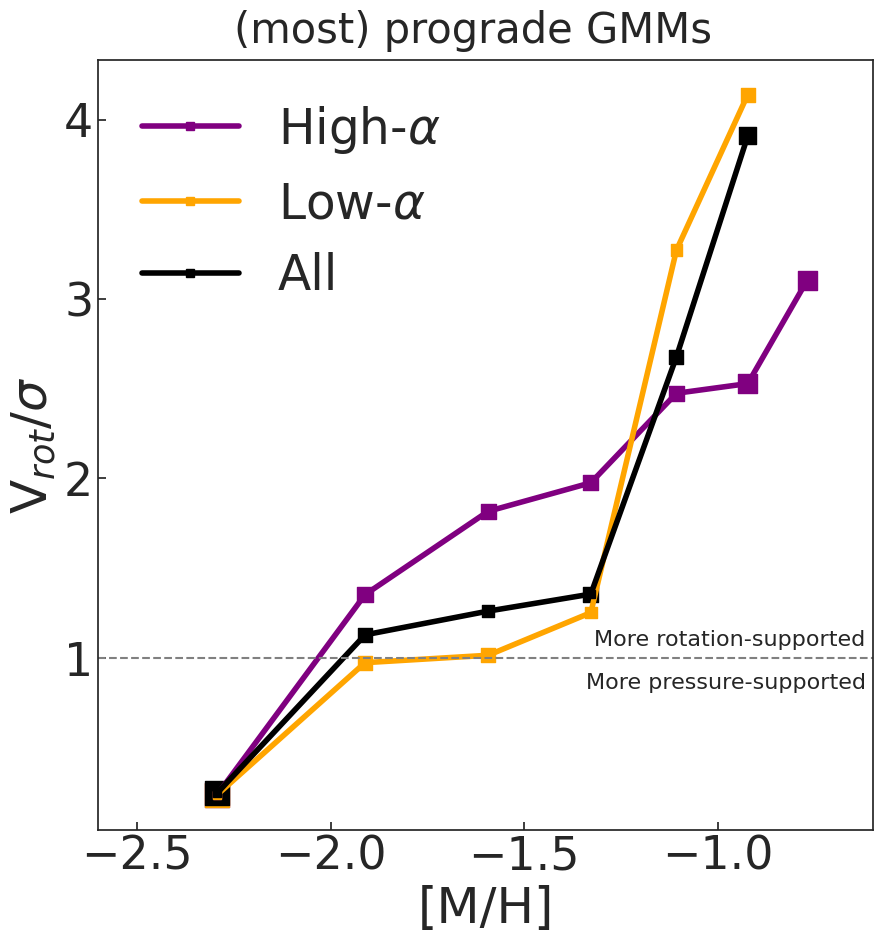}
    \caption{V$_\phi$/$\sigma_z$ ratio (azimuthal velocity to vertical velocity dispersion) for the most prograde Gaussian component in each metallicity bin (spaced evenly on a log scale) from our GMM runs in Section \ref{4.2}. The square size indicates the relative weight of the most prograde Gaussian in each metallicity bin. This ratio is a unitless quantity indicating the level of rotational support within a disc, essentially quantifying how dynamical cold it is at $z=0$. The dashed line at V$_\phi$/$\sigma_z = 1$ marks the boundary between pressure-supported (halo-like) vs rotation-dominated kinematics (disc-like).}
    \label{fig:gmm-cam}
\end{figure}

In this subsection, we perform the traditional $N$-component GMM decomposition of high-$\alpha$, low-$\alpha$, and \textit{all stars} subsamples in bins of [M/H] on the 3D velocity space in cyclindrical coordinates -- v$_r$, v$_\phi$, and v$_z$. We set out to perform this exercise to understand how this method compares to our method presented in Section~\ref{3.3.1}. 

Typically, within any GMM framework, each sub-population is characterised by three key parameters. 
The first is a weighting factor, which reflects the relative proportion of stars belonging to that particular group. 
The second is the mean velocity, representing the first moment of the velocity distribution, and the average velocity along each of the three dimensions for stars within that group. 
Finally, the second moment of the velocity, the velocity dispersion, captures the variation in velocities around the average for stars within that group.
We use the same method outlined in \citet{2023zhang} for the GMM decomposition, to be able to compare the effects of $\alpha$-separation in this type of GMM decomposition.
To perform the GMM analysis, we utilised a software package named \texttt{pyGMMis} \citep{2016melchior}. 
This software incorporates the sophisticated "Extreme Deconvolution" technique developed by \citet{2011bovy} to account for the inherent uncertainties in stellar velocity measurements and the covariances between the input parameters.

Our analysis focuses specifically on metal-poor stars. We restricted the GMM decomposition to stars with a metallicity range of --2.5 to --0.7, expressed as [M/H]. Below this we are hampered by low number statistics and above these metallicities we are affected by the non-Gaussian nature of the low-$\alpha$ and high-$\alpha$ discs distribution in velocity space that could compromise the accuracy of the GMM results. 
The [M/H] bins are spaced equidistantly on a log scale with seven(six) bins for the high-(low-)$\alpha$ samples. 
The \textit{all stars} sample is fit using the same six bins as in the low-$\alpha$ sample.\footnote{For the \textit{all stars} and low-$\alpha$ stars samples, we neglect the most metal-rich bin as it is affected by the disc's asymmetry and suggests up to 8 components as the optimal fit, which is un-physical.} \citet{2023zhang} excluded stars located further than $|z|>2.5$ kpc from the Galactic plane, which is a cut that we do not use to be able to study the entire underlying data. We argue that as we are modelling the halo and disc populations separately, this cut is unnecessary. 
For each metallicity bin, we applied the GMM to the three-dimensional velocity space defined by the cylindrical galactocentric coordinates: radial velocity (v$_r$), azimuthal velocity (v$_\phi$), and vertical velocity (v$_z$). 
Moreover, each star's measurement uncertainty was incorporated as a covariance matrix with the diagonals representing the variances for each velocity component and the off-diagonals representing the covariance between a pair of velocities. To determine the optimal number of Gaussian components at each metallicity bin, we employed a standard approach based on the Bayesian Information Criterion (BIC), defined in the same way as  \citet{2023zhang}. Here, lower BIC values indicate a better fit, balancing model complexity with data fidelity.
To mitigate the risk of getting stuck in a local minima during the optimisation (sub-optimal solutions), we repeated the GMM fitting process 100 times with different starting conditions and recorded the BIC value for each trial.
The fit with the lowest BIC value (in the median BIC curve) was considered the optimal solution, suggesting a high likelihood of finding the globally best fit.
Increasing the complexity of the model by adding an extra component, while keeping the log-likelihood unchanged, raises the BIC value by $\sim100$. 
Given this order of magnitude, some $N$-component GMMs have very similar BIC values, especially in relatively metal-rich regimes, where the rotating component is fit with multiple Gaussians instead of 1. 
When BIC values are similar, we prefer models with fewer components, as they offer a more straightforward physical interpretation. 
This occurs for at least 1 bin in high-$\alpha$ and 2 bins in low-$\alpha$ and \textit{all stars} sub-samples.
We do not compute the uncertainty in the GMM parameters as \citet{2023zhang} found them to be generally around $\sim0.1$ km s$^{-1}$. Thus, uncertainties in the fits can be treated as negligible.

In Figure \ref{fig:gmm-cam}, we show the ratio of azimuthal velocity to vertical velocity dispersion (V$_\phi$/$\sigma_{{z}}$), which is commonly used as a measure of how rotation-supported or 'disc-like' a sample of stars is.
We restrict our analysis to the most prograde GMM component (largest mean v$_\phi$) per [M/H] bin to trace the evolution of the Milky Way's disc; these are also the component with an almost zero mean v$_r$ and v$_z$. In addition, to ensure that our model captures the halo and disc components well, we verified that the other GMM components the model finds fit for a halo-like substructure (V$_\phi$/$\sigma_z$ << 1). 
However, as we are only interested in the evolution of the disc-like GMM, we restrict our analysis to the most prograde GMM component in this work in Figure \ref{fig:gmm-cam}.

From Figure \ref{fig:gmm-cam}, we can see that the high-$\alpha$ stars gradually increase in their V$_\phi$/$\sigma_{{z}}$ (in purple) towards higher metallicity bins while low-$\alpha$ (in orange) and \textit{all stars} (in black) have a more rapid increase in V$_\phi$/$\sigma_{z}$ between two [M/H] bins, $\sim-1.3$ and $\sim-1$. 
The size of the scatter points are directly proportional to their relative weights in each bin.
Therefore, our results suggest that not accounting for the $\alpha$-separation could make the spin-up seem more rapid and exponential over a small range of metallicities. In our case, as we have been able to distinguish high-/low-$\alpha$ populations, we find that the high-$\alpha$ prograde GMMs favour a gradual spin-up of a metal-poor high-$\alpha$ population (likely the proto-Galaxy) to a metal-rich high-$\alpha$ disc over increasing metallicities. 
The GMMs from \citet{2023zhang} show a two-component fit in the most metal-poor bin while our model fits only one component.
This is due to the fact that our \textit{all stars} sample is composed of all stars with reliable metallicities from \citet{2023aandrae}, while having an $\alpha$ estimate from \citet{2024li}, whereas \citet{2023zhang} only used the metallicity estimates and therefore, we end up having far fewer VMP stars than theirs. 
\citet{2023zhang}'s results also suggest a more rapid growth in V$_\phi$/$\sigma_{\phi}$ over a shorter range in metallicity that is slightly more metal-poor than ours ($-1.7<$ [M/H] $<-1.3$), which is explained by the difference in bin sizes and the scale height $|z|$ < 2.5 kpc cut. 
GMM decomposition can be highly dependent on the choice of the [M/H] bins. 
However, this GMM decomposition is performed in order to make a qualitative comparison between high- and low-$\alpha$ samples, and is not a one-to-one comparison to the literature results. 
The evolution of the velocity components and its dispersion focused on the high-$\alpha$ stars tracing the evolution of the proto-Galaxy is shown in Appendix \ref{C}.

In summary, our results reveal that the high-$\alpha$ subsample shows a gradual rise in V$_\phi$/$\sigma_{{z}}$, favouring a gradual spin-up phase for the Milky Way's high-$\alpha$ disc over increasing metallicities.

\subsection{Limitations and future scope}\label{4.4}

In this work, we use an $\alpha$-separation on \textsl{Gaia} XP based [$\alpha$/M] and [M/H] abundances to separate high-/low-$\alpha$ stars and use the high-$\alpha$ stars to trace the azimuthal velocity (v$_\phi$) evolution of the old Milky Way disc over a range of metallicities ($-2.5<\rm [M/H]<0.1$).
The high-$\alpha$ selection implemented in this work is a simple piece-wise function (as described by equations \ref{eq:1} and \ref{eq:2}), and is supposed to select only in situ stars.
However, at lower metallicities [M/H]<--1.5, accreted merger populations (e.g. \textsl{Heracles}, the high-alpha tail of the GES and possibly other merger remnants) overlap, so a simple $\alpha$-cut can no longer establish a clean separation between in situ versus accreted population purely. 
This is modelled with our mixture model with evolving mean velocities and velocity dispersion for the spin-up phase and a background halo model that is isotropic as described in section \ref{3.3.1} and \ref{4.1}.
However, this model assumes a Gaussian distribution of azimuthal velocity at any given metallicity, which is not strictly true for the high-$\alpha$ disc, which has an asymmetric drift (long tail towards lower v$_{\phi}$, see Figure \ref{fig:circ-main}).
Because of this, the model tries to inflate the halo velocity dispersion at higher metallicities to fit this tail of the high-$\alpha$ disc which is nonphysical.
This also makes the model yield strong systematic residuals when compared to the data in the metal-rich end, where the high-$\alpha$ disc displays a strong prograde profile.
This is a consequence of a simple Gaussian assumption that the model is based on.
In the future, it would be possible to extend this model by using 
a more sophisticated dynamically motivated distribution function that represents more accurately the Galaxy's disc/halo in order to fully measure the evolution of velocities from the chaotic proto-Galactic state to an ordered high-$\alpha$ disc state with respect to metallicities. 
In doing so, we could also use the total velocity distributions to constrain the enclosed mass and in turn, measure the mass growth of the Milky Way from a proto-Galactic state at high-redshift to the old high-$\alpha$ disc state as a function of increasing metallicity.
We reserve this exploration for a future work.

It is also important to note that the measured mean velocities and velocity dispersions with respect to decreasing metallicities are all present-day velocities and not the velocity they had at formation. 
From cosmological simulations, there have been clues that the Galaxy must have undergone post-formation dynamical heating which could change the net rotation that we see at present-day \citep{2024mccluskey,2024horta}. 

One other factor to keep in mind is the ability to precisely capture the subtle trends in [$\alpha$/Fe] with our high-$\alpha$ sample, given our $\alpha$ separation. 
Recent studies such as \citet{2022belokurov} and \citet{2022conroy} used \textsl{APOGEE} and \textsl{H3} surveys, respectively, to show that the [$\alpha$/Fe] ratio gradually declines between $-3<$ [Fe/H] $<-1.3$, and then rises to a higher value to meet the high-$\alpha$ disc population.
It is very likely that we are not catching this dip in [$\alpha$/Fe] with our simple $\alpha$-separation.
However, this should not affect the main conclusions of this work. 

It is also important to recognise that the high-[$\alpha$/Fe] disc population is not a single-age cohort, but rather spans a range of formation times and chemical enrichment levels. In particular, stars with the highest [$\alpha$/Fe] values (e.g. $\geq$ 0.3 dex) likely formed very early—within the first few hundred million years of Galactic evolution—while others in the high-$\alpha$ sequence formed later, albeit still early on Galactic timescales \citep{1990matteucci,1997chiappini,2019hasselquist}. While our analysis focuses on the median azimuthal velocity and orbital circularity evolution of the high-$\alpha$ disc as a whole, disentangling how internal structure in [$\alpha$/Fe] within this population contributes to the observed trends is an avenue for future exploration.

\section{Conclusions and outlook}\label{5}

In this work we set out to model the azimuthal velocity evolution of high- and low-$\alpha$ stars across metallicity space using \textsl{Gaia} XP element abundances, DR3 astrometry, and RVS radial velocities. By employing various mixture models, we have uncovered several lines of evidence that provide new insights into the kinematics of the metal-poor high-$\alpha$ disc and its evolution within the Milky Way.

Our analysis reveals that the metal-poor high-$\alpha$ disc shows a gradual and monotonic increase in its average azimuthal velocity distribution over a broad range of [M/H], spanning approximately $-1.7 <$ [M/H] $< -1$. This finding supports the scenario of a gradual spin-up of the metal-poor high-$\alpha$ disc, likely representing the proto-Galaxy, evolving into a rotationally supported high-$\alpha$ disc as [M/H] increases. This gradual transition underscores the dynamic evolution of the early Milky Way over a large range in metallicities, reflecting a continuous and progressive change in the rotational characteristics of the metal-poor high-$\alpha$ stellar population.

In contrast, the low-$\alpha$ sample presents a different picture due to the superposition of the GES and other accreted debris along with the low-$\alpha$ disc. The transition from metal-poor (halo) populations to metal-rich (disc) populations in this sample is much sharper, resembling a step-function at [M/H] $\sim -1$. The dominance of the GES debris in the metal-poor sample for all stars in our dataset results in a similar profile when inspecting the \textsl{Gaia} XP sample without any $\alpha$-selection. This sharp transition highlights the distinct formation and evolutionary histories of these stellar populations compared to the more gradual evolution in rotation across metallicities observed in the high-$\alpha$ disc.

Our results emphasise the critical importance of [$\alpha$/M] selection for studying the azimuthal velocity (or the orbital circularity) evolution of the old Milky Way disc. By distinguishing between high- and low-$\alpha$ populations, we can better understand the complex interplay between different stellar populations and their contributions to the overall dynamics of the Galaxy. This distinction helps us to disentangle the effects of accreted debris and in situ star formation, thus providing a clearer picture of the Milky Way's formation history and its subsequent evolution.
However, we note that this separation, while useful, does not lead to a purely in situ population.

In conclusion, our study provides strong evidence for the gradual spin-up of the metal-poor high-$\alpha$ disc, suggesting a continuous and progressive evolution from the proto-Galaxy to a rotationally supported high-$\alpha$ disc over increasing metallicities, which has been recently reported in \citet{2024bhorta} using \textit{APOGEE-Gaia} data's detailed chemical abundances and velocity distributions. The sharp transition observed in the low-$\alpha$ sample underscores the impact of accreted material on the Galaxy's dynamical structure. These findings contribute to our understanding of the early evolutionary processes of the Milky Way and highlight the need for detailed chemical and kinematic studies to unravel the complex history of our Galaxy. With the advent of more sophisticated distribution functions and increased availability of abundances for a larger number of stars, we  be better equipped to perform detailed analyses of in situ stars, further unraveling the intricate history of the Milky Way and refining our understanding of its earliest formation and evolution.

\begin{acknowledgements}
We express our gratitude to the reviewer for dedicating their valuable time and providing insightful comments that significantly enhanced the quality of our manuscript. 
This work was developed for the CCA Pre-Doctoral Program in 2023 at the Flatiron Institute. We thank them for their generous support.
AV thanks Vasily Belokurov for suggesting the test using high-resolution spectroscopic surveys. 
AV also thanks Ewoud Wempe for the helpful discussions on Gaussian processes, Tom Callingham for the helpful discussions on orbital circularity, Amina Helmi for the useful comments on this work, that helped improve our model, and Alis Deason for the insightful conversations on this topic.
AV gratefully acknowledges support from the Canadian Institute for Theoretical Astrophysics (CITA) through a CITA National Fellowship and the International Astronomical Union (IAU) and the Gruber Foundation through a IAU Gruber Fellowship.
DH was supported by the UKRI Science and Technology Facilities Council under project 101148371 as a Marie Curie Research Fellowship.
ES acknowledges funding through VIDI grant "Pushing Galactic Archaeology to its limits" (with project number VI.Vidi.193.093) which is funded by the Dutch Research Council (NWO).
This work has been partially supported by a Spinoza Prize from NWO.
This work has made use of data from the European Space Agency (ESA) mission \textsl{Gaia} (https://www.cosmos.esa.int/gaia), processed by the \textsl{Gaia} Data Processing and Analysis Consortium (DPAC, https://www.cosmos.esa.int/web/gaia/dpac/consortium). Funding for the DPAC has been provided by national institutions, in particular the institutions participating in the \textsl{Gaia} Multilateral Agreement.
AV also thanks the availability of the following packages and tools that made this work possible: \texttt{vaex} \citep{2018vaex}, \texttt{pandas} \citep{2022pandas}, \texttt{astropy} \citep{2022astropy}, \texttt{NumPy} \citep{2006numpy,2011numpy}, \texttt{SciPy} \citep{2001scipy}, \texttt{matplotlib} \citep{2007matplotlib}, \texttt{seaborn} \citep{2016seaborn}, \texttt{agama} \citep{2019vasiliev}, \texttt{gala} \citep{2017pricewhelan}, \texttt{galpy} \citep{2016jupyter}, \texttt{healpy} \citep{healpy}, \texttt{gaiadr3-zeropoint} \citep{2021lindegreen}, \texttt{jax} \citep{2019jax}, \texttt{numpyro} \citep{2019phan}, \texttt{scikit-learn} \citep{2011scikit-learn}, \texttt{pyGMMis} \citep{2016melchior}, \texttt{JupyterLab} \citep{2016jupyter}, and \texttt{topcat} \citep{2018topcat}.
\end{acknowledgements}

%
%

\bibliographystyle{mnras}
\bibliography{aanda}

\begin{appendix} 

\section{[M/H]-v$_\phi$ plane towards and away from the inner Galaxy}\label{A}

\begin{figure*}
    \centering
    \includegraphics[width=\textwidth]{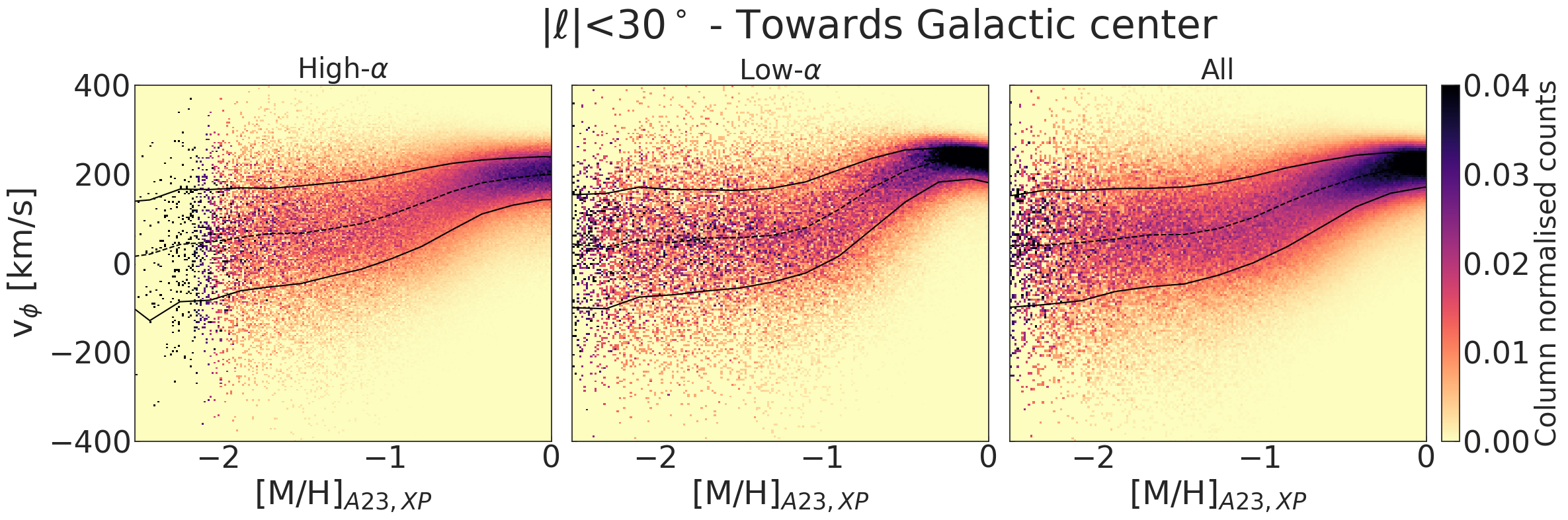}
    \includegraphics[width=\textwidth]{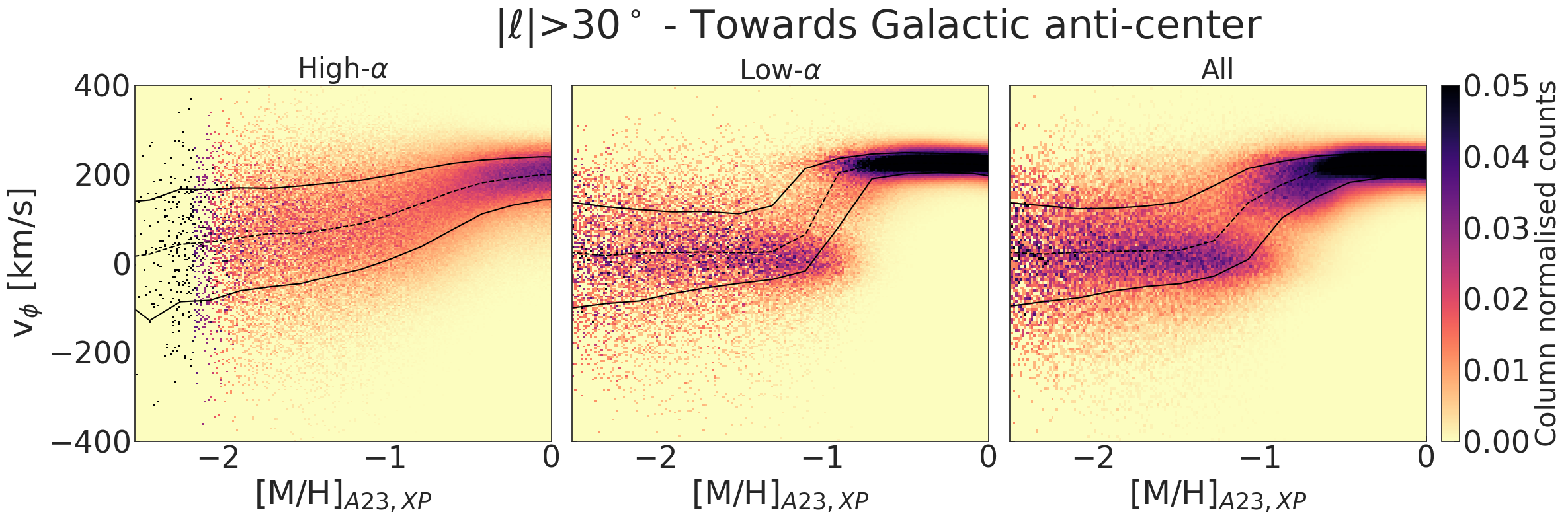}
    \caption{Column-normalised (by sum) 2D histogram of stars in the [M/H]-v$_\phi$ plane for all the stars (right), high-$\alpha$ selection (left), and low-$\alpha$ selection (middle) towards the Galactic centre (top panels) and away from the Galactic centre (anti-centre, bottom panels). The running median track is shown as dashed black line and the 16${th}$ and 84${th}$ percentile tracks are shown as black lines in all panels. We can see that the low-$\alpha$ stars are not fully two separate populations towards the inner Galaxy due to contamination from the proto-Galaxy in the low-$\alpha$ end, that creates a small connection between the thin disc and halo populations in the overall [M/H]-v$_\phi$ plane for low-$\alpha$ stars as seen in Figure \ref{fig:median-vphi-feh}. Low-$\alpha$ stars towards the Galactic anti-centre show a cleaner, two separate population of thin disc and halo stars as expected. This effect is also reflected in the \textit{all stars} panel. This effect is minimal in the \textsl{APOGEE} $\alpha$-selection, most likely due to the high-resolution $\alpha$-measurements.}
    \label{fig:cen-anticen}
\end{figure*}

\begin{figure}
    \centering
    \includegraphics[width=\columnwidth]{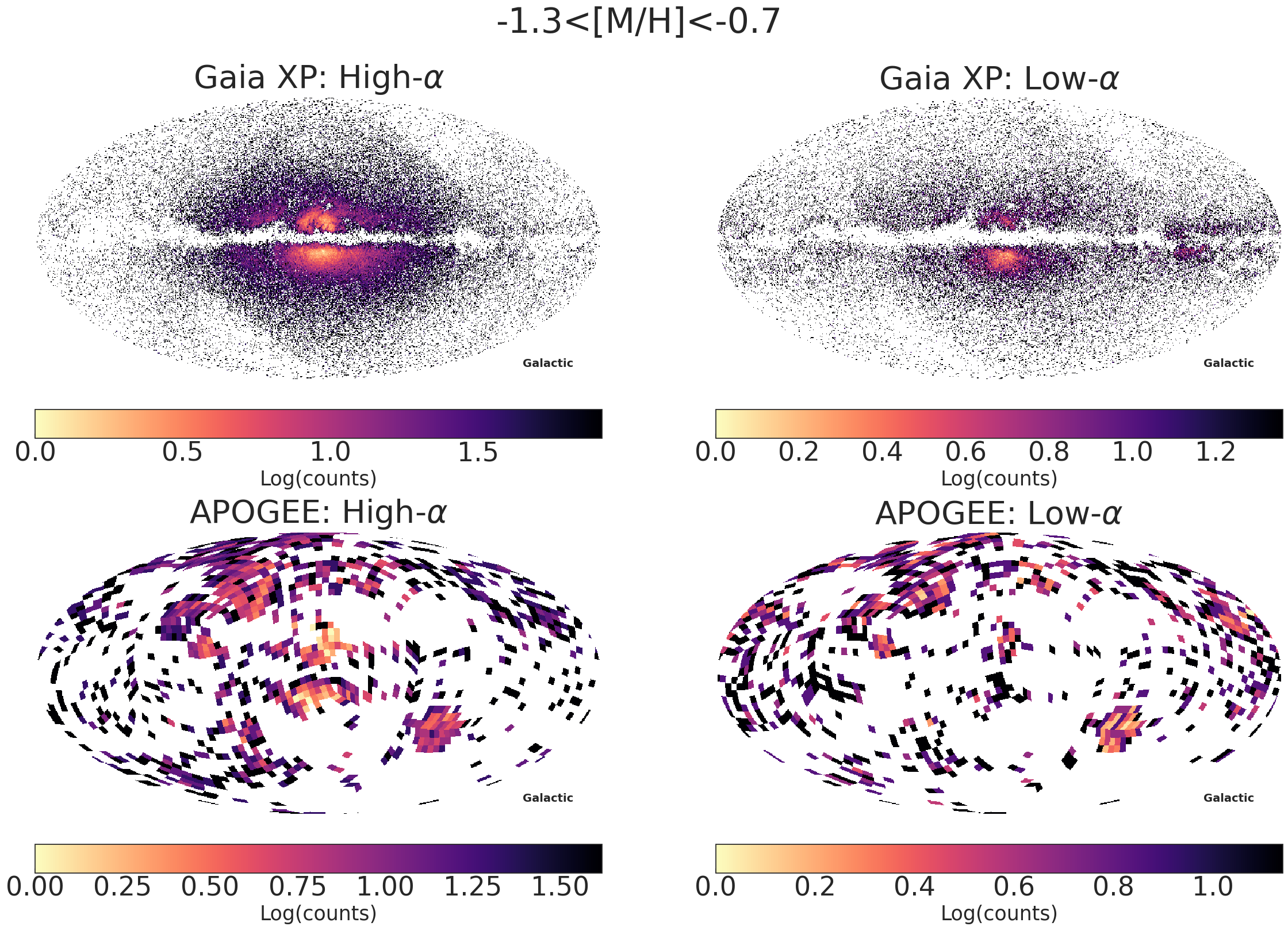}
    \caption{Logarithmic Galactic map of our sample of stars with high-$\alpha$ (left) and low-$\alpha$ (right) selections for the all-sky \textsl{Gaia} XP sample (top) and \textsl{APOGEE} DR17 sample (bottom), in the metallicity regime where the gradual spin-up in high-$\alpha$ vs the step function due to two different stellar populations in higher and lower metallicities in low-$\alpha$ is reflected the most: --1.3<[M/H]<--0.7. In the Gaia XP sample, we can see that the poor old heart or the proto-Galaxy is still present (contamination) in low-$\alpha$ towards the inner Galaxy, however, to a smaller extent than in high-$\alpha$. This is not the case for \textsl{APOGEE} stars which have a much cleaner low-$\alpha$ population towards the inner Galaxy with the bulk of proto-Galaxy towards the inner Galaxy in the high-$\alpha$ Galactic map.}
    \label{fig:on-sky-cen-anticen}
\end{figure}

Examining the [M/H]-v$_\phi$ plane for low-$\alpha$ stars in the \textsl{APOGEE} (Figure \ref{fig:median-apogee}) data, we can see that the halo-like (isotropic) population at low metallicity is almost fully disconnected from the higher metallicity disc-like component. However, this feature is not as immediately clear when inspecting the same stellar populations in the \textsl{Gaia} XP (Figure \ref{fig:median-vphi-feh}) data (there appears to be more of a connection between the halo-like and disc-like population for low-$\alpha$ stars). In this Appendix, we set out to investigate if this difference is due to the sample selection difference between \textsl{APOGEE} and \textsl{Gaia}, that could lead to more high-$\alpha$ stars contaminating our low-$\alpha$ star sample, by looking at the difference between the [M/H]-v$_\phi$ plane towards and away from the Galactic centre.

One possible reason for the discrepancy is that proto-Galactic fragments tend to be more centrally concentrated \citep{2024horta}. As \textsl{APOGEE} is a near-infrared survey that can penetrate through prevalent dust extinction in the central regions of the Galaxy more easily, it is likely that we are probing the proto-Galaxy better with \textsl{APOGEE} than with the (optical) \textsl{Gaia} survey. Thus, if our $\alpha$ cut was designed well, the low-$\alpha$ star sample should not be as centrally concentrated as the high-$\alpha$ population, that hosts both more centrally concentrated (high-$\alpha$) disc stars \citep{2015hayden, 2023Imig} and proto-Galactic populations \citep{2024horta}.

Figure \ref{fig:cen-anticen} shows the [M/H]-v$_\phi$ plane for all, low-$\alpha$, and high-$\alpha$ stars towards Galactic longitudes in the direction of the Milky Way Centre (i.e. |$\ell|<30^\circ$, top panels), and towards Galactic longitudes away from the Galactic Centre (i.e. the Galactic anti-centre, |$\ell|>30^\circ$, bottom panels) for our \textsl{Gaia} XP sample.
For low-$\alpha$ stars, we can see that the halo-like population at lower metallicities is fully disjoint from the disc-like population at higher metallicities for stars towards the Galactic anti-centre (similar to what is seen in \textsl{APOGEE}, Figure \ref{fig:median-apogee}), whereas the transition in v$_\phi$ across different metallicities is smoother towards the Galactic centre, similar to high-$\alpha$ median tracks. We reason that this is because, when looking towards the Galactic anti-centre, the two dominant populations contributing to the low-$\alpha$ regime are: 1) the (outer) low-$\alpha$ disc that is highly rotating; 2) the debris from the GES merger, which are highly radial. This also affects the \textit{all stars} panel for stars towards the Galactic centre, as seen in the top panels of Figure \ref{fig:cen-anticen}.

This leads us to infer that there exists a slowly rotating in situ proto-Galaxy-like population contaminating our low-$\alpha$ sample that is centrally concentrated, which is likely not present in the \textsl{APOGEE} stars.
This arises mainly due to the higher abundance precision in \textsl{APOGEE} (high-resolution spectroscopic survey) compared to \textsl{Gaia}-XP inferred $\alpha$-abundances.

In Figure \ref{fig:on-sky-cen-anticen}, we show the on-sky distribution of stars in Galactic coordinates in \textsl{Gaia} XP sample with a healpix level of 7 (top panels) and \textsl{APOGEE} sample with a healpix level of 4 (bottom panels) for both high- and low-$\alpha$ selection between the metallicities of --0.7 and --1.3. 
We choose this metallicity range because this is the range between which we see the difference in the v$_\phi$ median tracks between low-$\alpha$ stars towards and away from the Galactic centre as shown in Figure \ref{fig:cen-anticen}. 
In both \textsl{APOGEE} and \textsl{Gaia} XP high-$\alpha$ stars, we see centrally concentrated distribution of metal-poor stars, reminiscent of the `poor old heart' \citep{2022rix} also known as the proto-Galactic in situ population.
However, we also see a higher density of stars towards the inner Galaxy in the \textsl{Gaia} XP low-$\alpha$ panel, which is not as clear an overdensity in the \textsl{APOGEE} low-$\alpha$ panel. 
Within the inner Galaxy (|$\ell$|<30$^\circ$ and |$b$|<30$^\circ$), the high-$\alpha$ stars are overdense compared to the low-$\alpha$ stars by a factor of 8 for \textsl{APOGEE} sample while the high-$\alpha$ stars are overdense compared to the low-$\alpha$ stars by a factor of 4 for the \textit{Gaia} XP sample. 
Therefore, this mismatch between \textsl{APOGEE} and \textit{Gaia} XP is not just due to lower number statistics in \textsl{APOGEE}.
This difference could be due to unreliable $\alpha$ estimates and simple definition of $\alpha$-separation.
Therefore, in situ versus accreted separation using \textsl{Gaia} XP $\alpha$ estimates is not as reliable and the low-$\alpha$ selection using \textit{Gaia}-XP sample is more contaminated than the \textsl{APOGEE} sample, which is most likely the reason for a shallower step function in the low-$\alpha$ v$_\phi$ median tracks in \textsl{Gaia} XP (bottom left panel of Figure \ref{fig:median-vphi-feh}) compared to the steeper one in \textsl{APOGEE} (middle panel of Figure \ref{fig:median-apogee}).
However, we do not expect our conclusions to be affected by this contamination as our high-$\alpha$ (mostly in situ) selection is still pure. 

\section{Comparison of orbital circularity versus metallicity with \citet{2023chandra} results}\label{B}

\begin{figure*}
    \centering
    \includegraphics[width=\textwidth]{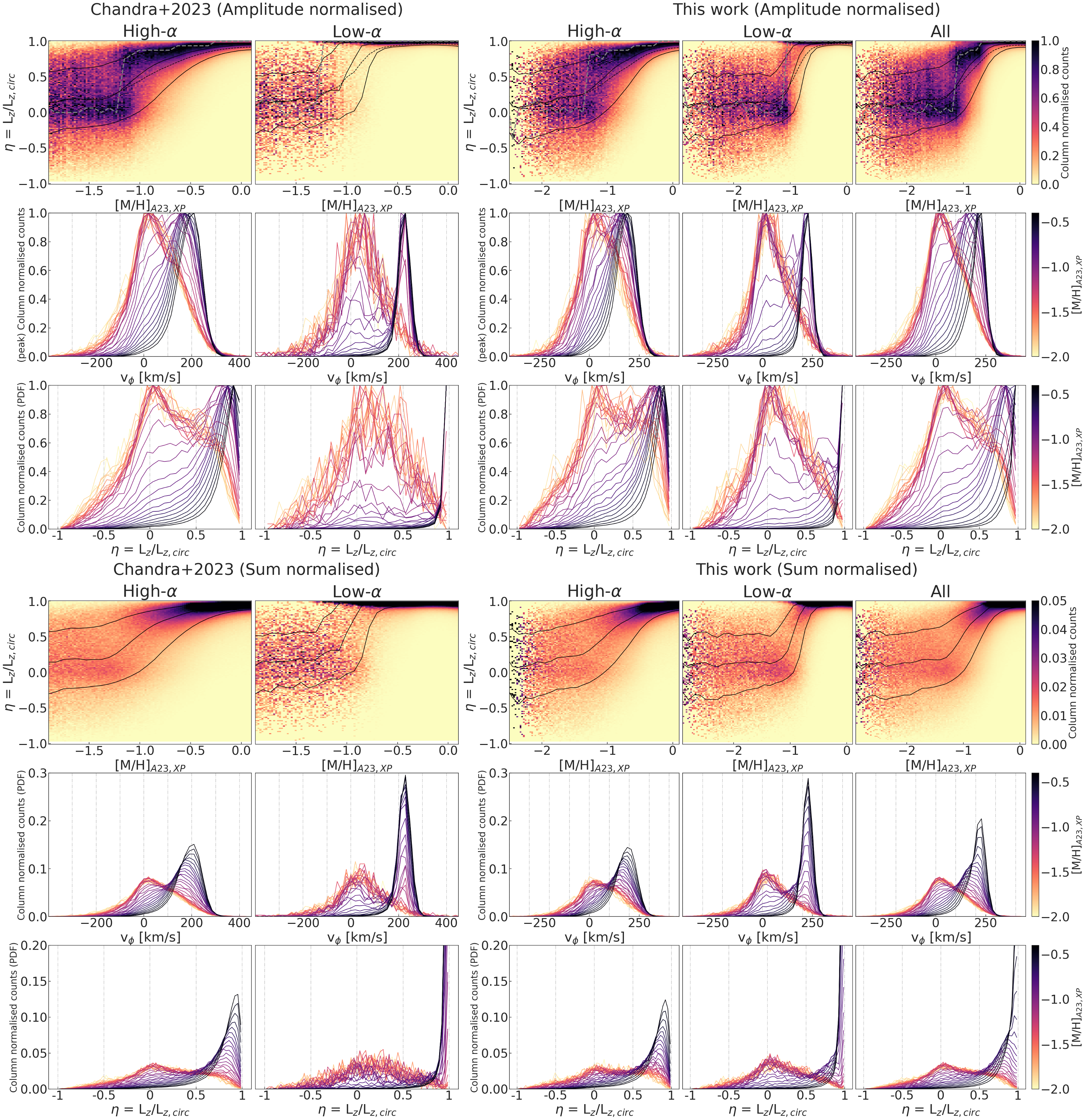}
    \caption{Orbital circularity vs metallicity for high-$\alpha$, low-$\alpha$, and \textit{all stars}, and 1D histogram of azimuthal velocity and orbital circularity at different metallicity bins for column-normalisation by amplitude (top panels) and by sum (bottom panels) using \citet{2023chandra} $\alpha$-selection (left panels) vs the $\alpha$-selection implemented in this work (right panels).
    Running median (black line), mean (black dashed line), mode (grey dashed line), 16${th}$ and 84${th}$ percentile (black lines) tracks are overlaid for the 2D histograms with column normalisation by amplitude. We can see that the mean and median tracks follow each other closely except at higher metallicities where the non-Gaussianity of thin and thick discs dominates, whereas the mode tracks follow the peaks of the background and is very noisy and resembles the step function behaviour seen by \citet{2023chandra}, even for high-$\alpha$ stars where we see a gradual spin-up (across increasing [M/H]) using column normalisation by sum. Running median, 16${th}$ and 84${th}$ percentile tracks are overlaid as black lines for the 2D histograms with column normalisation by sum. We also see that our $\alpha$-selection has a cleaner high-$\alpha$ sample and isolates the bulk of GES into the low-$\alpha$ sample.}
    \label{fig:comp-chandra}
\end{figure*}

\begin{figure}
    \centering
    \includegraphics[width=\columnwidth]{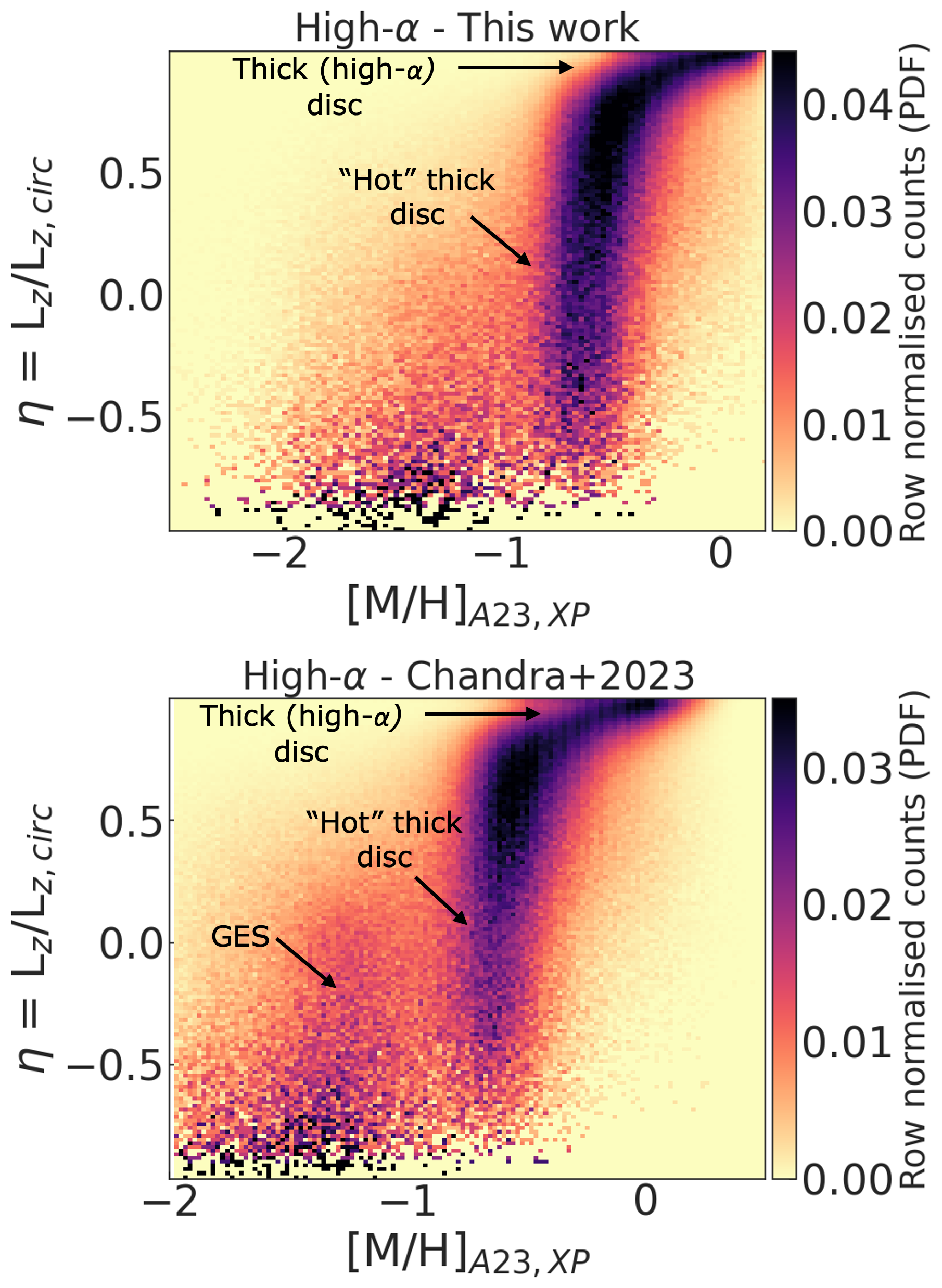}
    \caption{Row-normalised conditional metallicity distribution for high-$\alpha$ stars across orbital circularities for our $\alpha$-selection (top) and \citet{2023chandra} $\alpha$-selection (bottom). The high-$\alpha$ disc, in situ halo (hot high-$\alpha$ disc), and accreted components can be seen in both panels. However, the contamination from the accreted GES merger is stronger in the \citet{2023chandra} $\alpha$-selection than the $\alpha$-selection proposed in this work.}
    \label{fig:row-norm}
\end{figure}

In this section, we compare the evolution of orbital circularity versus metallicity in our work with the results from \citet{2023chandra}.
Both the works use the same input catalogue based on \textsl{Gaia} XP spectra.
The main differences between our approaches are the difference in the $\alpha$-separation, and the way the column normalisation is performed.
In Figure \ref{fig:comp-chandra} top panels, we compare the column-normalised by amplitude 2D histograms of orbital circularity versus metallicity and their corresponding $\eta$, and v$_\phi$ 1D histograms for \citet{2023chandra} $\alpha$-selection and the $\alpha$-selection described in this work (equations \ref{eq:1} and \ref{eq:2}).
In all the panels, the 16${th}$, 50${th}$, and 84${th}$ percentile tracks are shown as black lines, mean tracks are shown as black dashed lines and mode tracks are shown as grey dashed lines.
It is important to note that the mode tracks trace the underlying distribution the best when the column normalisation is calculated by the amplitude of the distribution.
In Figure \ref{fig:comp-chandra} bottom panels, we compare the column-normalised by sum 2D histograms of orbital circularity versus metallicity and their corresponding $\eta$, and v$_\phi$ 1D histograms for \citet{2023chandra} $\alpha$-selection and the $\alpha$-selection described in this work (equations \ref{eq:1} and \ref{eq:2}).
This is equivalent to what is presented in the rest of this paper (tracing the PDF of each distribution, equivalent to normalising such that the area under the curve is equal to 1).
In all the panels, the 16${th}$, 50${th}$, and 84${th}$ percentile tracks are shown as black lines.

There are two main inferences that can be made from this comparison, as discussed below:
\begin{itemize}
    \item The $\alpha$-selection described in this work is more efficient in removing the last major merger (accreted GES) from the high-$\alpha$ selection (in situ equivalent), better than the $\alpha$-selection described by \citet{2023chandra}. Therefore, our high-$\alpha$ stars are cleaner than the high-$\alpha$ stars from \citet{2023chandra}. 
    \item Column normalisation of 2D histograms can be done in many ways. 
    From our comparison, we conclude that the column normalisation is scaled by the amplitude of the distribution in \citet{2023chandra}, whereas in this work, we column normalise the histograms by the sum of the distribution. 
    Column normalisation by amplitude scales the peak of each 1D histogram to 1.0 whereas column normalisation by sum scales the sum of the histogram (equivalent to integral under the histogram curve) to be equal to 1.0 thereby tracing the probability distribution function of each 1D histogram. 
    Column normalisation by amplitude traces the peak of each 1D histogram with no easily interpretable connection between each 1D histograms, and therefore produces a noisy view of the underlying data. 
    This can be seen by the noisy streaks in the 2D histograms and the noisy mode tracks in the top panels of Figure \ref{fig:comp-chandra}.
    Column normalisation by amplitude traces the mode of the distribution, which is also tricky in case of bimodal distributions. The metal-poor end of our underlying azimuthal velocity is bimodal due to a slowly rotating proto-Galactic population and the high-$\alpha$ remnants of accreted stellar systems (mostly isotropic). 
    The mode simply traces the peak of the distribution and therefore looks like a step function when one Gaussian dominates over the other in relative weights. 
    This affects the high-$\alpha$ population the most as the low-$\alpha$ stars are already two almost disjoint distributions in metallicities (thin disc at higher metallicities and accreted halo at lower metallicities). 
    Therefore, we emphasise that column normalisation by sum is more appropriate to understand the underlying distribution of orbital circularity or azimuthal velocity as it traces the PDF of the distribution and not just the peak/the mode. 
\end{itemize}

These two reasons together explain why one would interpret the spin up to be rapid and drastic, whereas the underlying distribution is slowly gaining rotation over a wide range of metallicities.

In order to better understand the difference in the $\alpha$-separation between our work and \citet{2023chandra}, we also show row normalised 2D histograms of [M/H]-$\eta$ space for high-$\alpha$ stars in this work and high-$\alpha$ stars from \citet{2023chandra} in the top and bottom panels of Figure \ref{fig:row-norm}.
Because of row-normalisation, higher metallicity stars are more highlighted in these figures.
In both these panels, we can see the highly rotating high-$\alpha$ disc dominating higher metallicities ([M/H]>--0.4). 
Below these metallicities, we see high-$\alpha$ stars with a broad range of circularities, even down to retrograde orbits, but have metallicities that are representative of high-$\alpha$ disc. 
The most plausible origin for these stars are that they we born in the old high-$\alpha$ disc and got kicked-up into halo-like orbits by the last major merger, GES \citep{2017bonaca,2018helmi,2020belokurov}.
The most interesting difference between our high-$\alpha$ stars and that of \citet{2023chandra} is that their high-$\alpha$ stars have larger number of stars in isotropic orbits around [M/H]$\sim$--1.2, reminiscent of the GES merger. 
This is almost absent in our high-$\alpha$ stars.
This leads us to believe that our high-$\alpha$ star selection is purer than the one in \citet{2023chandra}.
It is easier for the eye to trace the excess of retrograde low-metallicity stars in both the panels, but these are only highlighted due to the row-normalisation (because retrograde stars are almost only present due to halo accretion events in the lower-metallicity end) and in reality, there are far fewer of these stars.
However, we still see a population of accreted stars in our high-$\alpha$ selection (to a much lower extent than using \citet{2023chandra} $\alpha$-separation), which can be attributed to the evolution of any stellar system that has a high-$\alpha$ low-metallicity tail which cannot simply be removed using a simple $\alpha$-separation.
We model this population along with the evolution of high-$\alpha$ disc in subsection \ref{3.3.1} and subsection \ref{4.1}.

\section{Velocity evolution versus metallicity for high-$\alpha$ stars used as a cosmic clock}\label{C}

\begin{figure}
    \centering
    \includegraphics[width=\columnwidth]{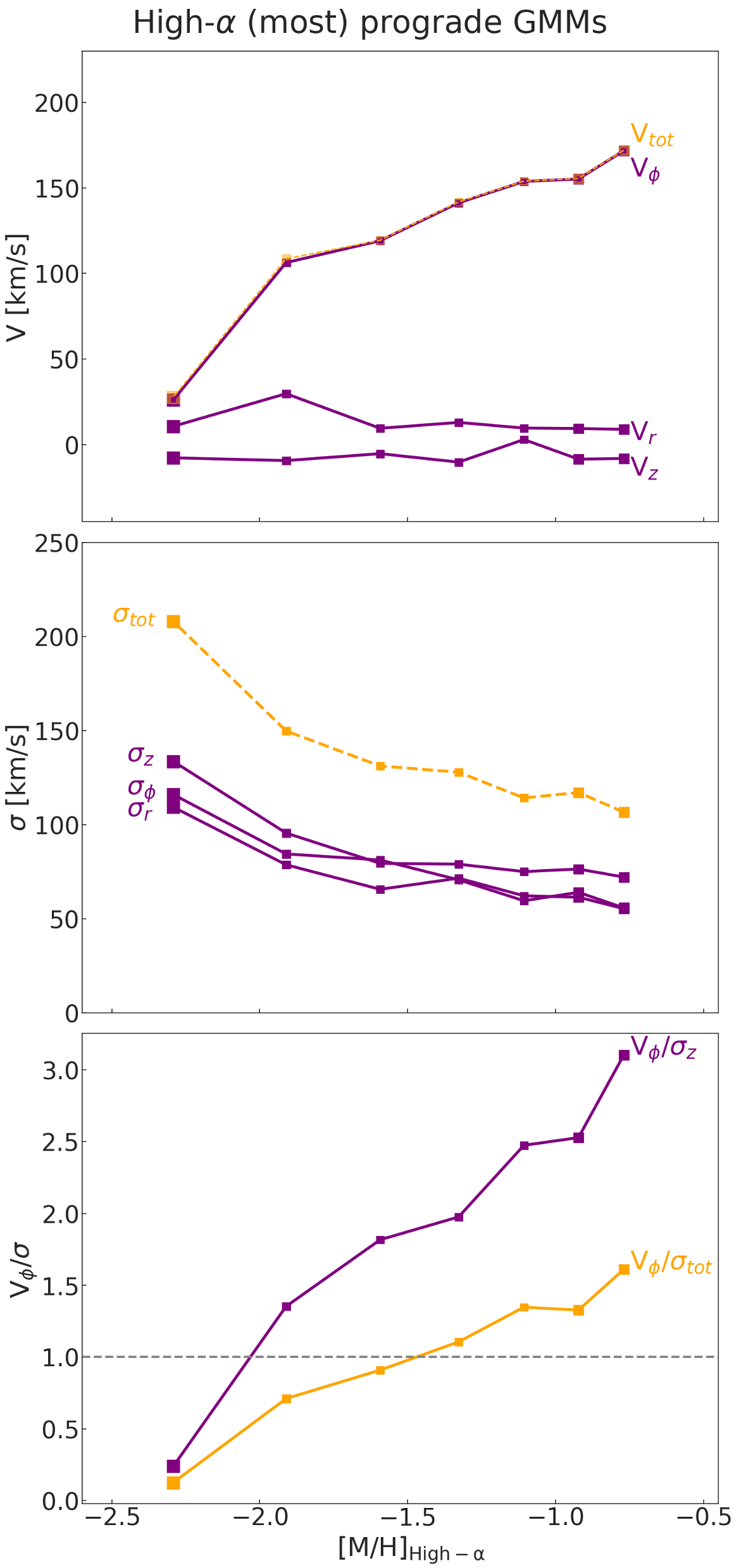}
    \caption{Mean velocities (top), velocity dispersions (middle) and V$_\phi$/$\sigma_z$ ratio (bottom) showing how rotationally supported the stars are, for the most prograde Gaussian component in each metallicity bin from our GMM runs for high-$\alpha$ stars.}
    \label{fig:v-phi-normal-gmm}
\end{figure}

In this section, we show the evolution of the most prograde GMM component (based on the GMM runs explained in subsection \ref{4.2}) in its different velocity components for high-$\alpha$ stars. 
In Figure \ref{fig:v-phi-normal-gmm}, we see the evolution of (V$_r$, V$_\phi$, V$_z$) in the top panel, the evolution in their velocity dispersion ($\sigma_r$, $\sigma_\phi$, $\sigma_z$) in the middle panel and their rotational support (V$_\phi$/$\sigma_z$ or V$_\phi$/$\sigma_{tot}$) in the bottom panel across different [M/H] bins with the respective component names next to each curve.
It is important to note that these velocities and their dispersions trace the present dynamical evolution and not those at formation.
Using FIRE-2 simulations, \citet{2024mccluskey} shows that the rotational velocities can increase now compared to at formation in the pre-disc era due to stars being torqued into rotational orbits as the disc settles, and the rotational velocities can decrease now compared to at formation in the late-disc era due to dynamical heating. In case of rotational velocity dispersion, it also does not directly reflect the formation history, as it monotonically increases due to post-formation dynamical heating, adding to the velocity dispersion at formation.  
Therefore, even though metal-poor stars trace old stellar populations, the velocities do not simply trace the formation velocities.
However, this simple analysis of velocity evolution can give us an idea of the evolution of the high-$\alpha$ disc as we see it now. 

In the top panel of Figure \ref{fig:v-phi-normal-gmm}, we see that the radial velocity and vertical velocity is almost close to zero, with the rotational velocity (and the total velocity) increasing slowly with increasing metallicities. 
This is reminiscent of a slowly rotating proto-Galactic population that settles into a high-$\alpha$ disc at higher metallicities. 
We see that this spinning-up phase is more gradual over increasing metallicities than previously reported.
However, if this gradual spin-up comes from the time of its formation or due to post-formation heating is an open question.
In the middle panel of Figure \ref{fig:v-phi-normal-gmm}, we see the trends for velocity dispersions in all three directions decreasing with increasing metallicities, as the high-$\alpha$ disc begin to settle. 
In the bottom panel of Figure \ref{fig:v-phi-normal-gmm}, we see the evolution of rotational support as a function of metallicity.
\citet{2024mccluskey} show that in late- and early-disc era, both v$_\phi$/$\sigma_z$ and v$_\phi$/$\sigma_{tot}$ decrease by a factor of 2 between formation and present state, due to post-formation dynamical heating that increases their velocity dispersions. 
In the bottom panel of Figure \ref{fig:v-phi-normal-gmm}, we see the evolution of rotational support gradually increasing with increasing metallicity, also reminiscent of a proto-Galactic population slowly gaining rotation and settling into the high-$\alpha$ disc. 
The metallicity at which this population becomes more rotation supported ('discy') is difficult to infer, as it differ between $\sigma_z$ and $\sigma_{tot}$, and also that the overall V$_\phi$/$\sigma_{tot}$ reduces at present when compared to what it was at formation. 
Therefore, our main conclusion from these velocity evolution curves is that the spin-up phase of a slowly rotating proto-Galaxy is gradual across a large range of metallicities and not as rapid as previously reported.

\end{appendix}
\end{document}